\documentclass[aps,prl,floatfix,twocolumn]{revtex4-1}
\usepackage{graphicx}
\usepackage{epstopdf}
\usepackage{dcolumn}
\usepackage{bm}
\usepackage{setspace}
\usepackage{psfrag}
\usepackage{float} 

\usepackage{graphicx}
\usepackage{epstopdf}
\usepackage{dcolumn}
\usepackage{bm}
\usepackage{setspace}
\usepackage{psfrag}
\usepackage{float} 

\usepackage[T1]{fontenc}
\usepackage[french]{babel}
\usepackage{chemfig}
\usepackage{hyperref}
\usepackage{fancyhdr}
\usepackage{lastpage}
\usepackage{graphicx}
\usepackage{pgf,tikz}
\usetikzlibrary{arrows}
\usepackage{amsmath}

\newcommand{\eeq}{\end{eqnarray}}

\begin{document}
\title{Phase-Resolved Two-Dimensional Spectroscopy of Electronic Wavepackets by Laser-Induced XUV Free Induction Decay}

\author{S. Beaulieu$^{1,2}$,
E. Bloch$^{1}$,
L. Barreau$^{3}$,
A. Comby$^{1}$,
D. Descamps$^{1}$,
R. G\'eneaux$^{3}$
F. L\'egar\'e $^{2}$,
S. Petit$^{1}$,
Y. Mairesse$^{1}$} 
\bigskip

\affiliation{
$^1$Universit\'e de Bordeaux - CNRS - CEA, CELIA, UMR5107, F33405 Talence, France\\
$^2$ ALLS, Institut National de la Recherche Scientifique, Centre EMT, J3X1S2, Varennes, Quebec, Canada\\
$^3$ LIDYL, CEA, CNRS, Universit\'e Paris-Saclay, CEA Saclay 91191 Gif-sur-Yvette, France\\
}

\date{\today}

\begin{abstract}

We present a novel time- and phase-resolved, background-free scheme to study the extreme ultraviolet dipole emission of a bound electronic wavepacket, without the use of any extreme ultraviolet exciting pulse. Using  multiphoton transitions, we populate a superposition of quantum states which coherently emit extreme ultraviolet radiation through free induction decay. This emission is probed and controlled, both in amplitude and phase, by a time-delayed infrared femtosecond pulse. We directly measure the laser-induced dephasing of the emission by using a simple heterodyne detection scheme based on two-source interferometry. This technique provides rich information about the interplay between the laser field and the Coulombic potential on the excited electron dynamics. Its background-free nature enables us to use a large range of gas pressures and to reveal the influence of collisions in the relaxation process. 
\end{abstract}
\maketitle

Transient Absorption Spectroscopy (TAS) in the extreme ultraviolet (XUV) range is a powerful technique for ultrafast dynamical studies, from the gas phase \cite{goulielmakis10,wang10,holler11} to the solid state \cite{schultze14,lucchini16}. The recent progress of attosecond science has made the extension of TAS to the attosecond regime (ATAS) a reality. In the past few years, different configurations of ATAS experiments have emerged. In the first and most intuitive scheme, the XUV attosecond pulses probe a sample that has been pre-excited by an infrared (IR) pump laser pulse. This scheme was for instance used to follow the ultrafast coherent hole dynamics initiated in strong-field ionized krypton \cite{goulielmakis10}. In the second arrangement, the XUV and IR pulses are temporally overlapped. The XUV absorption thus probes the IR-dressed atomic or molecular states, allowing the observation of light-induced states \cite{chen12,reduzzi15} as well as sub-cycle AC-Stark-shifts \cite{chini12}. Finally, in what turned out to be the most widely used scheme, the XUV pulse comes first and serves as a pump, exciting a broadband superposition of quantum states through single-photon absorption. The wavepacket relaxes by coherently emitting XUV radiation (called XUV Free Induction Decay, XFID) that interferes with the incident light. A delayed IR laser field is used to follow or control the relaxation dynamics, such that these experiments can be seen as Transient Reshaping of the Absorption spectrum of the XUV light (TRAX) \cite{beck14,cao16}.

\begin{figure}
\begin{center}
\includegraphics[width=0.48\textwidth]{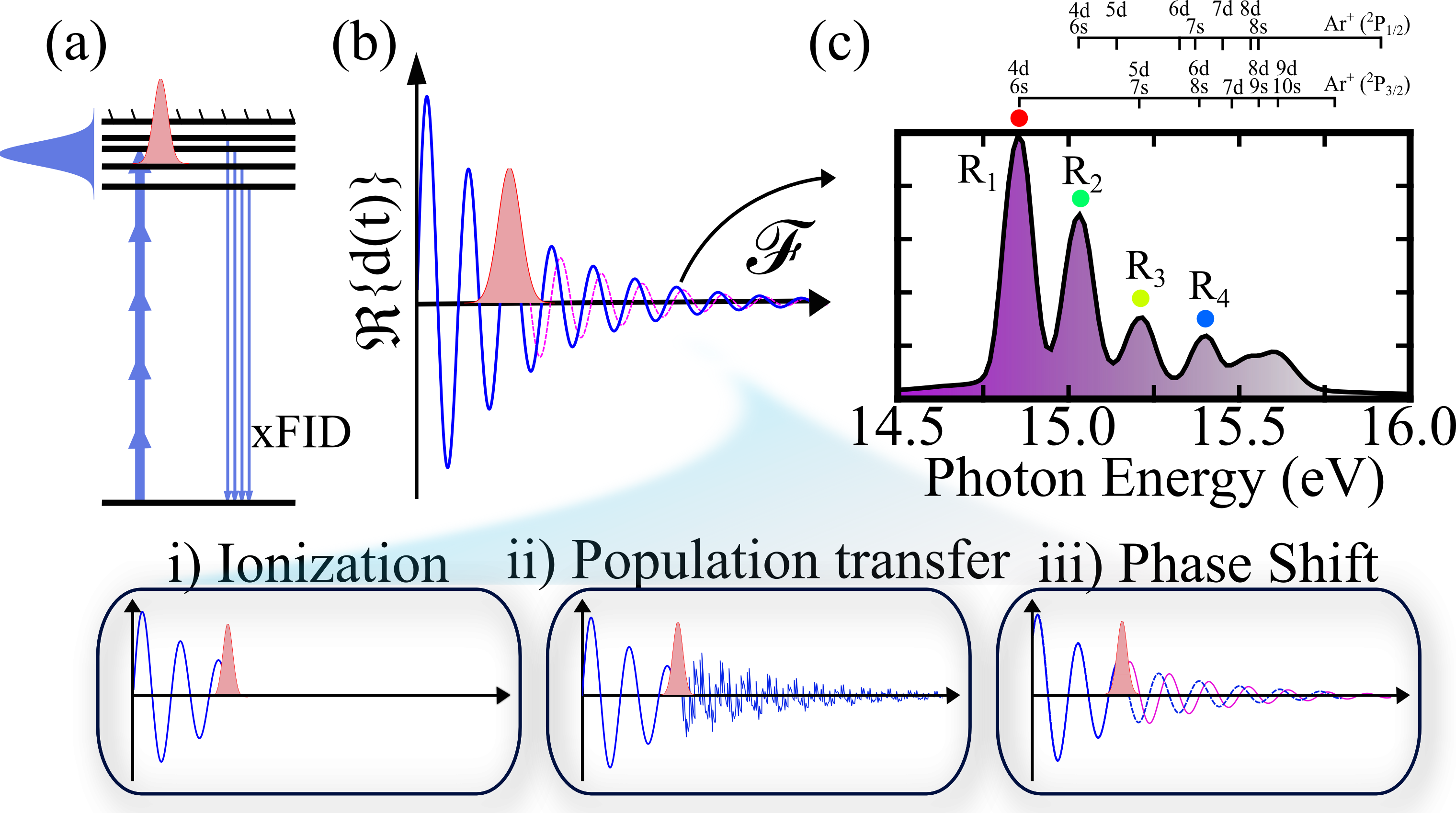}
\caption{Principle of the experiment: (a) Multiphoton excitation creates an electronic wavepacket in argon atoms, which emits XUV radiation through XFID. A time-delayed 800 nm laser pulse perturbs the XFID emission. (b) Schematic view of real part of the unperturbed and IR-perturbed atomic dipole with the different effects induced by the IR pulse, i.e. ionization, population transfer (field coupling) and Stark-shift induced phase shift. (c) Experimental XFID spectrum, which correspond to the Fourier transform of the atomic dipole.}
\label{fig1}
\end{center}
\end{figure}

In TRAX experiments, the delayed IR laser pulse has three main effects on the XFID emission (Fig. \ref{fig1}):
(i) Damping of the emission by ionization of the excited states. This effect induces a lowering and a spectral broadening of the absorption features \cite{wang10}.
(ii) Field coupling of different electronic states, leading to population transfers. The resulting amplitude reshaping of the electronic wavepacket can be detected through temporal beatings in the absorption signal, as demonstrated in atoms \cite{beck14,cao16} and molecules \cite{warrick16,cheng16}.
(iii) Phase shift induced by the Stark-shift of the excited states. This laser-induced phase was shown to enable full control over the absorption lineshapes, from Lorentz to Fano profiles \cite{ott13}. This effect was also used to spatially deflect the XFID radiation by creating a Stark-shift gradient in the medium \cite{bengtsson16}.
The combination of these different physical effects and observables provides a very rich framework for ultrafast spectroscopy and strong-field physics, enabling for instance the complete reconstruction of a two-electron wavepackets \cite{ott14}.

Nevertheless, TRAX suffers from several limitations that restrict its generalization. First, the phase sensitivity, which is a key component of TRAX, is based on the interferometric nature of the detection, which records the coherent superposition of the incoming XUV and radiated XFID light. The phase extraction is thus indirect and requires a modeling of the laser-atom interaction. Furthermore, when the XUV absorption cross section is small, the contribution of the incoming XUV to the interference pattern is much stronger than the radiated XFID, which leads to poor signal-to-noise ratio. One appealing solution to this problem is to increase the pressure-length product of the absorbing medium. However, recent works have demonstrated that this caused modifications of the absorption lineshapes due to resonant effects in the XUV pulse propagation, which complicates the interpretation \cite{liao15}.  

In this letter, we overcome the limitations of TRAX experiments by measuring Transient Reshaping of the Emission spectrum of the XUV light (T-REX). We excite a bound electronic wavepacket in argon atoms by absorption of 5 visible (VIS) photons, which then emits XFID radiation \cite{beaulieu16}. The spectrum of the exciting light does not overlap anymore with that of the emitted radiation. This provides a background-free XFID measurement. We manipulate the emission using a delayed IR pulse and show that all the key features of TRAX experiments can be observed. Since the measurement is not interferometric anymore and is therefore insensitive to the phase of the emission we implement a two-source interferometry scheme \cite{bellini98} to directly measure the XFID phase. The high sensitivity of our background-free technique enables us to use a broad range of gas pressures, revealing a strong influence of collisions in the T-REX signal even at backing pressures as low as a few tens of mbars, in a range where many TRAX measurements were performed. 

We used the 800 nm Ti:Sa Aurore laser system at CELIA which delivers 7 mJ, 25 fs pulses at 1 kHz repetition rate. The beam was sent into a Mach-Zehnder interferometer. The high-energy arm (90$\%$) was frequency-doubled by a 200 $\mu m$ thick type-I BBO crystal ($\theta$=29.2$^{\circ}$, $\phi$=90$^{\circ}$). In the low-energy 800 nm arm (10$\%$), the laser power and the polarization state were varied by rotating a superachromatic half-wave plate in front of a broadband polarizer. The dispersion in each arm of the Mach Zehnder was adjusted using silica plates to achieve optimal durations of both the 400 nm and 800 nm pulses. The two arms were recombined using a dichroic mirror and focused by a f = 1.5 m lens into a 250 $\mu m$ thick effusive gas jet of argon. The pump intensity I$_{400 nm}$ was on the order of few $10^{12}W/cm^2$. The XUV emission was analyzed by a flat-field XUV spectrometer, consisting of a 1200 grooves/mm cylindrical grating and a set of dual microchannel plates coupled to a phosphor screen. A 12-bit cooled CCD camera recorded the spatially-resolved XUV spectrum. 

A typical XFID spectrum using only the pump pulse is shown in Fig. 1(c). Because of the $\Delta \ell = \pm 1$ selection rule for the XUV photon emission, only the \textit{ns-} and \textit{nd-} Rydberg series can decay radiatively to the \textit{p-} ground state. These Rydberg manifolds converge to two different spin-orbit coupled ionic cores, $Ar^+ (^2 P_{3/2})$ and $Ar^+ (^2 P_{1/2})$, with ionization potentials of $I_p[3/2]=15.76 eV$ and $I_p[1/2]=15.94 eV$, respectively. The lines observed in Fig. 1(c) are spectroscopically identified according to \cite{yoshino70}. However the limited resolution of our XUV spectrometer ($\sim$ 60 meV around 15 eV) prevents the unambiguous assignment of all the spin-orbit states at this stage, which would require sub-10-meV spectral resolution.

\hspace{1cm}
\begin{figure}
\begin{center}
\includegraphics[width=0.48\textwidth]{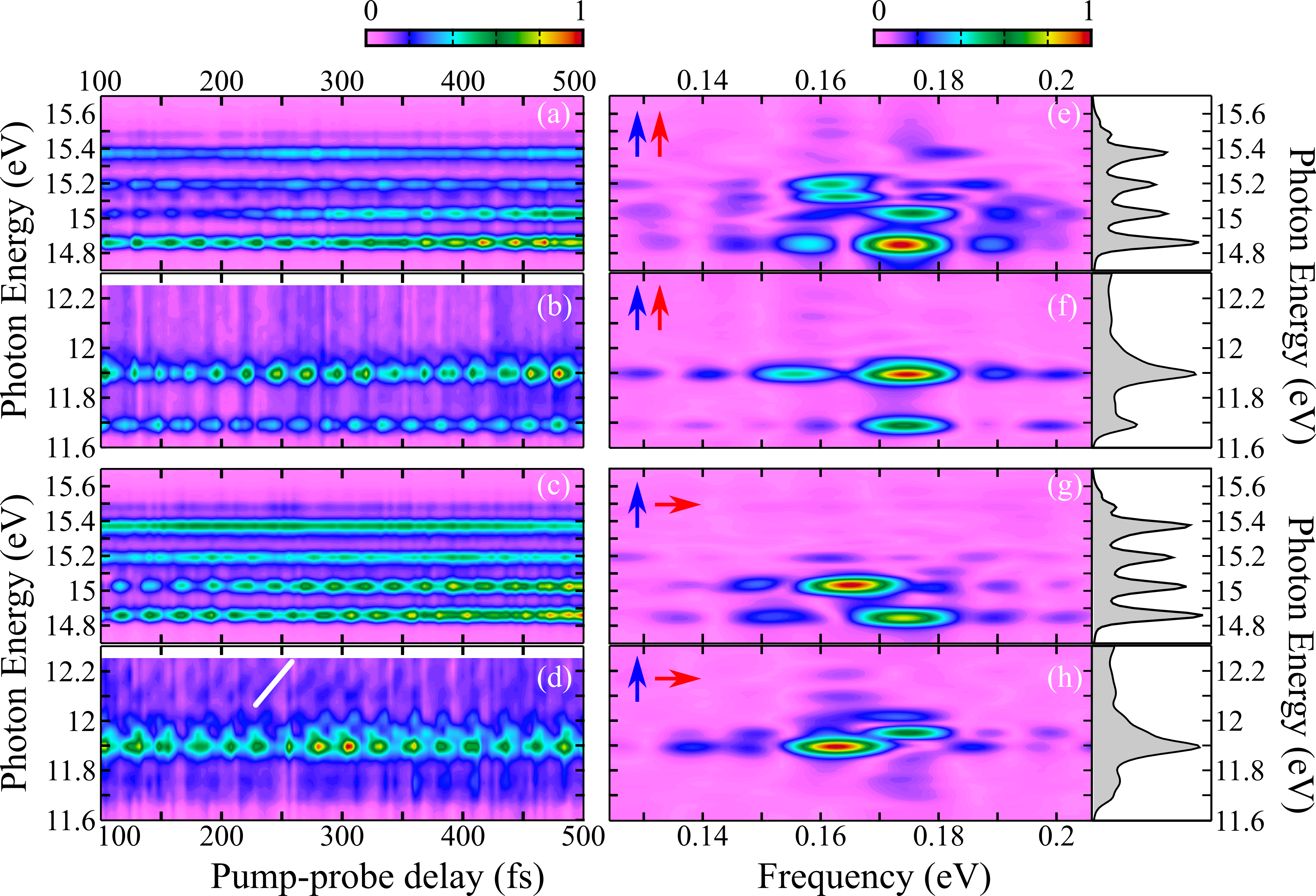}
\caption{Laser-induced population transfer in T-REX. Evolution of the Rydberg (a,c) and valence (b,d) emission spectrum as a function of pump-probe delay, in parallel (a,b) and perpendicular (c,d) pump-probe polarization configuration. (e-h) show the 2D spectra obtained by Fourier transform of each pump-probe scan. The white line in (d) serves as guide to the eyes to emphasis the linear dephasing of the beating with photon energy. The argon backing pressure was 5 mbar for both measurement and the probe intensity I$_{800 nm}$ $ \sim 2\cdot 10^{12}$W/cm$^{2}$. The cross-correlation time between the pump and the probe pulse is 70 fs.}
\label{fig2}
\end{center}
\end{figure} 

In order to investigate the coherence time of the XFID emission, we measured the time-dependent ionization damping of the XFID signal, using an IR probe pulse at $\sim 10^{13}$ W/cm$^{2}$. The XFID signal increases with delay, since the later the IR pulse arrives, the more XUV emission has occurred before ionization. The curves are shown in the SI. Exponential fittings provide rising times of 6.3 ps for R$_1$, 30.8 ps for R$_2$, 21.6 ps for R$_3$ and 33.6 ps for R$_4$. These are much shorter than the typical nanosecond lifetime of the Rydberg states, indicating the importance of homogeneous broadening in the measurement. The same measurement was repeated using different argon backing pressures. We observed a strong shortening of the XFID coherence time with increasing pressure: the coherence time was found to decrease by a factor 5 when the backing pressure increased from 10 to 1000 mbar. We will come back to this effect at the end of the paper. 

A finer pump-probe scan with 4 fs steps reveals fast oscillations of the XFID spectral components (Fig. 2(a-d)). These modulations are analyzed by Fourier-transforming each spectral slice to obtain 2D spectra (Fig. 2(e-h)), which give information about the coupling of each Rydberg state with the neighboring bound states. The spectral resolution of the measurement is set by the range of the delay scan and can reach the meV range, as illustrated in the SI. The signal from the two lower Rydberg states R$_1$ and R$_2$ shows a main Fourier peak at 0.174 eV, which corresponds to the frequency difference (0.1737 eV) between the 6s[3/2] and 6s[1/2] spin-orbit states. This indicates that these two states are mutually coupled by the IR field, in a two-photon $\Lambda$-type coupling. High resolution measurements (see SI) reveal additional modulations at 0.168 eV for these two peaks (population transfer between 4d[3/2] and 6s[1/2]), and an additional 0.155 eV modulation for the R$_2$ peak corresponding to the coupling between 4d[3/2] and 6s[3/2] (0.1552 eV). The emission from R$_3$ (15.19 eV) oscillates with a 0.164 eV frequency, corresponding to the difference between 6s[1/2] and 7s[3/2] (0.16341 eV). The highest peaks hardly show modulations, but a slow beating with a $\sim 200$ fs period  is visible in the R$_4$ (15.35 eV) signal. This beating must be caused by two oscillations that are too fast to be resolved in our experiments, but which are separated by $\sim 0.02$ eV energy. 

The XFID shows another spectral component between 11.6 and 12 eV, which is two orders of magnitude weaker and only exists for positive pump-probe delays. This indicates that it originates from laser-induced population transfer between Rydberg states and the unpopulated 4s states of argon (11.7229 eV and 11.8228 eV). Since the population at a given energy originates from several Rydberg states, we observe oscillations in the signal, corresponding to interferences between ladder type transitions \cite{cao16}. The oscillation frequencies are dictated by the energy differences between the Rydberg states involved, and are thus the same as in the $\Lambda$-type coupling modulating the Rydberg emission (0.164 eV and 0.174 eV measured in the high-resolution scan).  

The transition between the ground and the Rydberg states involve five VIS photons, and thus imposes a given orientation of the atomic dipole along the pump polarization axis. The fine pump-probe scan was repeated with perpendicular pump-probe relative polarization, and revealed different modes in the oscillating XFID signal; for instance, the dominant mode in the R$_2$ emission in the perpendicular case was the weakest in the parallel case (Fig. 2(g)). This shows that the population transfers are polarization-dependent. This observation is a signature of the angular dependence of the two-photon transition dipole matrix element. 

The effect of the relative polarization is even more dramatic in the 4s states spectral range. First, the emission around 11.7 eV almost disappears in the perpendicular polarization case. Moreover, as seen in Fig. 2(d), the emission spectrum is much more complex in the perpendicular case, showing broad structures whose different spectral components do not oscillate in phase. Surprisingly, the emission extends several hundreds of meV above the energy of the highest 4s state (11.8278 eV), and significantly below the first 4p (12.9070 eV). In order to understand how such broad spectral structures can emerge, we have to go beyond the impulsive approximation of the excitation process. The 800 nm laser pulses that transfer population from the 4p to the 4s states are intense enough to induced a Stark-shift of the states. Even though most of the XFID occurs after the probe IR pulse and is thus insensitive to this Stark-shift, a small fraction of the emission does occur during the probe pulse, i.e. from Stark-shifted states, creating a high-energy tail on the XFID line. Our data reveals that the phase of the oscillations increases quasi linearly within this tail, which is in good agreement with the laser-imposed phase model \cite{chen13}.  More generally, the investigation of the fast beatings shows the ability of T-REX to provide high-quality data for 2D spectroscopy. This scheme is conceptually similar to the XUV-IR four-wave mixing experiments recently demonstrated \cite{cao16-1}, and could be extended by using more sophisticated spatial and temporal combinations of excitation and probe pulses \cite{cao16-2}. 

\hspace{1cm}
\begin{figure}
\begin{center}
\includegraphics[width=0.48\textwidth]{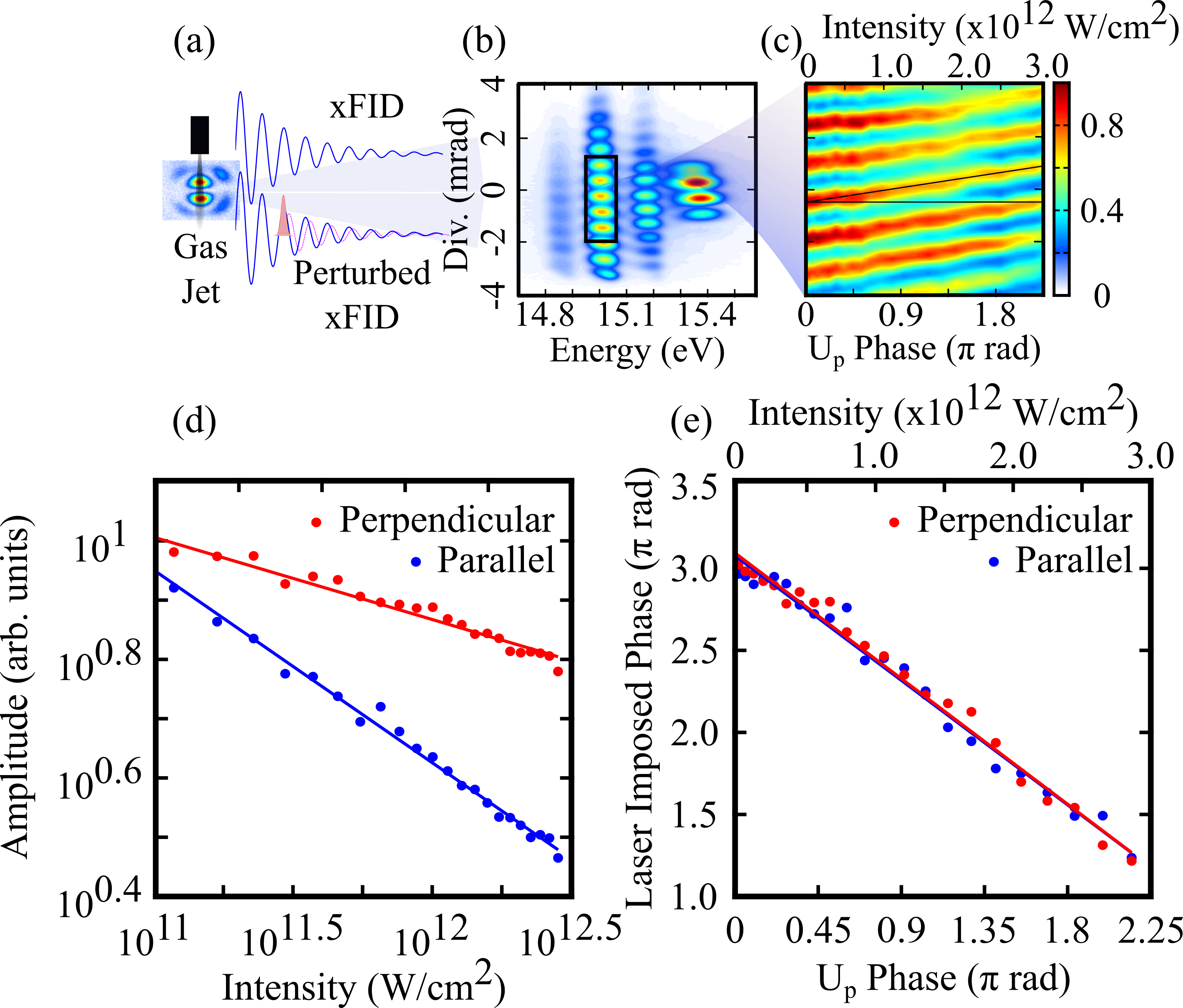}
\caption{(a) Schematic representation of the two-source interferometry experiment. The XFID emission in one of the sources is perturbed while the other one is used as a reference. (b) Spatially-resolved XFID spectrum. (c) Spatial profile of the R$_2$ XFID emission as a function of the IR pulse intensity at a pump-probe delay of 400 fs. (d) Amplitude, and (d) phase of the XFID emission of R$_2$ as a function of IR intensity, with parallel (blue) and perpendicular (red) pump-probe polarizations. The dots are the experimental points and the solid lines are linear fits. The argon backing pressure was 10 mbar.}
\label{fig3}
\end{center}
\end{figure}

One of the great advantages of TRAX experiments is their phase sensitivity. The transmitted part of the incident XUV pulse can be seen as the reference of an heterodyne detection system. The shape of the absorption lines acts as a quantum interferometer which enables to \textit{indirectly} measure the phase of the excited atomic dipole \cite{ott13}, with the help of theoretical models. We now show the possibility of \textit{directly} measuring this phase in a two-source T-REX measurement. We inserted a homemade, dry-etched fused-silica plate (see SI) which acts as a 0-$\pi$ phase mask in the 400 nm laser beam and split its focus in two spots \cite{camper15}. This creates two spatially separated XFID sources which interfere in the far field (Fig. 3(a)) \cite{bellini98}. Well-contrasted interference fringes were observed on the detector, reflecting the coherent nature of the XFID process. One of the sources was used as a reference, while the emission from the other was perturbed by an IR pulse. The fringe pattern was recorded as a function of the IR intensity, at a delay of 400 fs to avoid pump-probe overlap effects, and shows a linear shift of the fringes position (Fig. 3(a)). This reflects the dephasing of the emission induced by the IR pulse. A Fourier analysis was used to extract the XFID amplitude and phase (Fig. 3(b,c)). As the laser intensity increases, the signal decreases because of the damping by ionization. This damping is found to be much more efficient when the IR pulse polarization is parallel to the exciting pulse than when it is orthogonal. This means that it is easier to ionize the Rydberg states with a laser polarization parallel to the quantization axis. This observation, which is well known in tunnel ionization from valence orbitals \cite{young06,pavicic07,shafir09}, may seem surprising for such a low intensity. However, our experiment lies in an exotic regime where the $I_p$ of the excited atoms is less than 1 eV, such that the Keldysh parameter is close to 1. This proximity to the tunneling regime explains the strong dependence of the ionization yield on the polarization direction. This result shows that relatively weak laser fields can be used to transpose strong-field physics studies to highly excited states, opening the way to many applications in particular using mid-infrared laser sources. 

The dephasing of the perturbed XFID is determined by the time-integral of the AC-Stark-shift of the Rydberg states during the laser pulse. At the ionization threshold, the Stark-shift is equal to the ponderomotive energy ($U_p$) and the associated phase shift is $\phi_{U_p}$. As the electrons gets more bound, this shift is expected to decrease, because of their smaller polarizability. For the Rydberg states measured in our experiment we find slopes of 0.77$\cdot \phi_{U_p}$ for R$_4$, 0.72$\cdot \phi_{U_p}$ for R$_3$ and  0.66$\cdot \phi_{U_p}$ for R$_2$. Phase-resolved T-REX thus provides a direct evaluation of the AC-Stark-shifts, a quantity that is fundamentally important \cite{potvliege06} but not easy to accurately determine experimentally \cite{lopez-martens00}. The measured values are between 66 and 77 $\%$ of the ponderomotive shift, meaning that the Coulomb interaction between the low-lying Rydberg electrons and the ionic core is not negligible with respect to to the effect of the electric field of the laser on the electron dynamics. Moreover, our experiment shows that while the ionization yield is strongly affected by the relative polarization between pump and probe, the AC-Stark-shift is remarkably independent of this parameter. 

We have demonstrated that T-REX could provide accurate spectroscopic data on ultrafast laser-induced electronic wavepacket dynamics. On the other hand, the dynamics associated to the relaxation of the Rydberg states were found to be faster than expected, with a characteristic time decreasing with increasing backing pressure. To understand the origin of this discrepancy, we performed additional measurements of the pressure dependence of the XFID emission. First, we measured the XFID yield as a function of backing pressure over a very broad range (1-1600 mbar), and found a remarkably linear behavior. This is surprizing since a coherent emission is expected to scale quadratically with the number of emitters. A simple model derived in the SI shows that this is a direct effect of the atomic collisions which decrease the duration of the XFID emission as pressure increases. Thus, even if the argon density is rather low in the jet (at least one order of magnitude lower than the backing pressure), the delocalized nature of the Rydberg electrons causes significant collisional effects. In order to investigate the influence of collisions on the spectroscopic measurements, we repeated the femtosecond scans for different backing pressures (Fig. 4). As the pressure increases, the femtosecond oscillations of the XFID, which reflect the electronic wavepacket reshaping by the IR pulse, disappear, and a new spectral component, downshifting in energy as the pump-probe delay increases, appears.

\hspace{1cm}
\begin{figure}
\begin{center}
\includegraphics[width=0.48\textwidth]{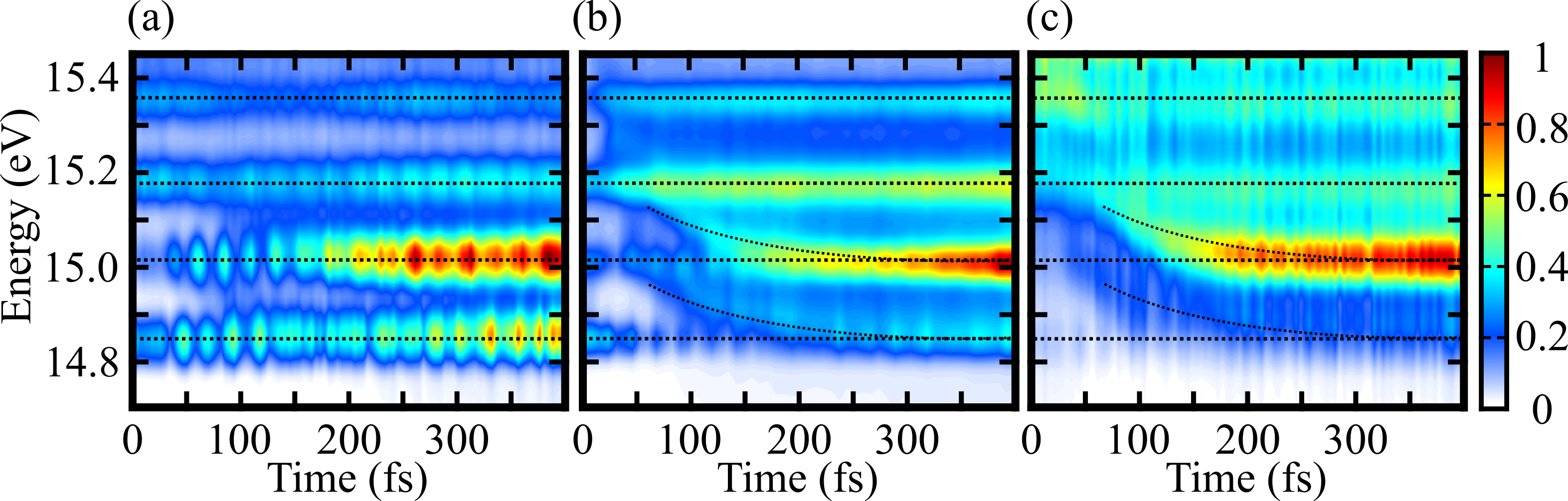}
\caption{Pressure effect in femtosecond T-REX scans. The XFID spectrum as a function of pump-probe delay at (a) 8 mbar, (b) 100 mbar and (c) 500 mbar argon backing pressure, with parallel pump-probe polarizations and I$_{800 nm}$ $ \sim 2\cdot 10^{12}$W/cm$^{2}$. The dashed black lines serves as a guide to the eyes.}
\label{fig2}
\end{center}
\end{figure}

To understand the effect of pressure in femtosecond T-REX scans, we have to account for the different contributions to the signal: the XFID emitted before the arrival of the probe pulse, which corresponds to the unperturbed dipole emission; the XFID emitted during the probe pulse, by field-dressed states; and the XFID emitted after the probe pulse, which contains signatures of the laser-induced reshaping of the electronic wavepacket (femtosecond beatings). At very low backing pressure (8 mbar, Fig 4(a)), the coherence time of the XFID is very long and most of the signal is emitted  after the probe pulse. Thus, the time-resolved spectrum is dominated by the beating caused by population transfers. At high backing pressure (500 mbar, fig 4(c)), collisions strongly reduce the XFID coherence time, which leads to the disappearance of the beating features and the appearance of a time-dependent energy shift of the spectral lines, which is signature of the Stark-shifts occuring during the probe pulse \cite{chini12}. At intermediate backing pressure (fig 4(b)), the spectrogram contains all these features, indicating a balance between the XFID signal emitted before, during and after the arrival of the NIR pulse. Pressure dependences have also been observed in absorption spectra and TRAX experiments. They have been interpreted as resulting from the reshaping of the propagating XUV exciting pulse by the strongly dispersive absorbing medium \cite{cao16-1}. While these effects certainly have an influence in TRAX, they cannot be at play in T-REX which is based on multiphoton excitation by VIS pulses.

In conclusion, we have demonstrated a novel background-free scheme to investigate the XUV dipole response of a bound electronic wavepackets. This technique directly provides both the amplitude and phase of the XUV free induction decay emission of the wavepacket, without the use of any XUV pulses. The laser-imposed phase measurement gives us an accurate determination of the Stark-shifts of these low-lying Rydberg states, providing values between 66$\%$ and 76$\%$ of $U_p$ for different states. The fact that we do not use any XUV pulse allows us to disentengle the effects of pulse propagation and collisions when increasing the atomic density, which cannot be done in standard XUV-Transient Absoption Spectroscopy experiments. Last, being based on optical pulses, T-REX is easily extendable to more complex configurations where coherences could be manipulated and revealed by multiple laser pulses, opening new dimensions in XUV spectroscopy. 

We thank V. Blanchet, F. Catoire, B. Fabre, B. Pons and R. Ta\"{i}eb for fruitful discussion, S. Guilet for designing and making the phase plate and R. Bouillaud and L. Merzeau for technical assistance. We acknowledge financial support of the French National Research Agency (ANR), through MISFITS (ANR-14-CE32-0014),  XSTASE (ANR-14-CE32-0010) and IdEx Bordeaux LAPHIA (ANR-10-IDEX-03-02). SB acknowledges the Vanier Scholarship. 

\bibliography{BiblioFromZotero.bib}

\begin{thebibliography}{28}%
\makeatletter
\providecommand \@ifxundefined [1]{%
 \@ifx{#1\undefined}
}%
\providecommand \@ifnum [1]{%
 \ifnum #1\expandafter \@firstoftwo
 \else \expandafter \@secondoftwo
 \fi
}%
\providecommand \@ifx [1]{%
 \ifx #1\expandafter \@firstoftwo
 \else \expandafter \@secondoftwo
 \fi
}%
\providecommand \natexlab [1]{#1}%
\providecommand \enquote  [1]{``#1''}%
\providecommand \bibnamefont  [1]{#1}%
\providecommand \bibfnamefont [1]{#1}%
\providecommand \citenamefont [1]{#1}%
\providecommand \href@noop [0]{\@secondoftwo}%
\providecommand \href [0]{\begingroup \@sanitize@url \@href}%
\providecommand \@href[1]{\@@startlink{#1}\@@href}%
\providecommand \@@href[1]{\endgroup#1\@@endlink}%
\providecommand \@sanitize@url [0]{\catcode `\\12\catcode `\$12\catcode
  `\&12\catcode `\#12\catcode `\^12\catcode `\_12\catcode `\%12\relax}%
\providecommand \@@startlink[1]{}%
\providecommand \@@endlink[0]{}%
\providecommand \url  [0]{\begingroup\@sanitize@url \@url }%
\providecommand \@url [1]{\endgroup\@href {#1}{\urlprefix }}%
\providecommand \urlprefix  [0]{URL }%
\providecommand \Eprint [0]{\href }%
\providecommand \doibase [0]{http://dx.doi.org/}%
\providecommand \selectlanguage [0]{\@gobble}%
\providecommand \bibinfo  [0]{\@secondoftwo}%
\providecommand \bibfield  [0]{\@secondoftwo}%
\providecommand \translation [1]{[#1]}%
\providecommand \BibitemOpen [0]{}%
\providecommand \bibitemStop [0]{}%
\providecommand \bibitemNoStop [0]{.\EOS\space}%
\providecommand \EOS [0]{\spacefactor3000\relax}%
\providecommand \BibitemShut  [1]{\csname bibitem#1\endcsname}%
\let\auto@bib@innerbib\@empty
\bibitem [{\citenamefont {Goulielmakis}\ \emph {et~al.}(2010)\citenamefont
  {Goulielmakis}, \citenamefont {Loh}, \citenamefont {Wirth}, \citenamefont
  {Santra}, \citenamefont {Rohringer}, \citenamefont {Yakovlev}, \citenamefont
  {Zherebtsov}, \citenamefont {Pfeifer}, \citenamefont {Azzeer}, \citenamefont
  {Kling}, \citenamefont {Leone},\ and\ \citenamefont
  {Krausz}}]{goulielmakis10}%
  \BibitemOpen
  \bibfield  {author} {\bibinfo {author} {\bibfnamefont {E.}~\bibnamefont
  {Goulielmakis}}, \bibinfo {author} {\bibfnamefont {Z.-H.}\ \bibnamefont
  {Loh}}, \bibinfo {author} {\bibfnamefont {A.}~\bibnamefont {Wirth}}, \bibinfo
  {author} {\bibfnamefont {R.}~\bibnamefont {Santra}}, \bibinfo {author}
  {\bibfnamefont {N.}~\bibnamefont {Rohringer}}, \bibinfo {author}
  {\bibfnamefont {V.~S.}\ \bibnamefont {Yakovlev}}, \bibinfo {author}
  {\bibfnamefont {S.}~\bibnamefont {Zherebtsov}}, \bibinfo {author}
  {\bibfnamefont {T.}~\bibnamefont {Pfeifer}}, \bibinfo {author} {\bibfnamefont
  {A.~M.}\ \bibnamefont {Azzeer}}, \bibinfo {author} {\bibfnamefont {M.~F.}\
  \bibnamefont {Kling}}, \bibinfo {author} {\bibfnamefont {S.~R.}\ \bibnamefont
  {Leone}}, \ and\ \bibinfo {author} {\bibfnamefont {F.}~\bibnamefont
  {Krausz}},\ }\href {\doibase 10.1038/nature09212} {\bibfield  {journal}
  {\bibinfo  {journal} {Nature}\ }\textbf {\bibinfo {volume} {466}},\ \bibinfo
  {pages} {739} (\bibinfo {year} {2010})}\BibitemShut {NoStop}%
\bibitem [{\citenamefont {Wang}\ \emph {et~al.}(2010)\citenamefont {Wang},
  \citenamefont {Chini}, \citenamefont {Chen}, \citenamefont {Zhang},
  \citenamefont {He}, \citenamefont {Cheng}, \citenamefont {Wu}, \citenamefont
  {Thumm},\ and\ \citenamefont {Chang}}]{wang10}%
  \BibitemOpen
  \bibfield  {author} {\bibinfo {author} {\bibfnamefont {H.}~\bibnamefont
  {Wang}}, \bibinfo {author} {\bibfnamefont {M.}~\bibnamefont {Chini}},
  \bibinfo {author} {\bibfnamefont {S.}~\bibnamefont {Chen}}, \bibinfo {author}
  {\bibfnamefont {C.-H.}\ \bibnamefont {Zhang}}, \bibinfo {author}
  {\bibfnamefont {F.}~\bibnamefont {He}}, \bibinfo {author} {\bibfnamefont
  {Y.}~\bibnamefont {Cheng}}, \bibinfo {author} {\bibfnamefont
  {Y.}~\bibnamefont {Wu}}, \bibinfo {author} {\bibfnamefont {U.}~\bibnamefont
  {Thumm}}, \ and\ \bibinfo {author} {\bibfnamefont {Z.}~\bibnamefont
  {Chang}},\ }\href {\doibase 10.1103/PhysRevLett.105.143002} {\bibfield
  {journal} {\bibinfo  {journal} {Physical Review Letters}\ }\textbf {\bibinfo
  {volume} {105}},\ \bibinfo {pages} {143002} (\bibinfo {year}
  {2010})}\BibitemShut {NoStop}%
\bibitem [{\citenamefont {Holler}\ \emph {et~al.}(2011)\citenamefont {Holler},
  \citenamefont {Schapper}, \citenamefont {Gallmann},\ and\ \citenamefont
  {Keller}}]{holler11}%
  \BibitemOpen
  \bibfield  {author} {\bibinfo {author} {\bibfnamefont {M.}~\bibnamefont
  {Holler}}, \bibinfo {author} {\bibfnamefont {F.}~\bibnamefont {Schapper}},
  \bibinfo {author} {\bibfnamefont {L.}~\bibnamefont {Gallmann}}, \ and\
  \bibinfo {author} {\bibfnamefont {U.}~\bibnamefont {Keller}},\ }\href
  {\doibase 10.1103/PhysRevLett.106.123601} {\bibfield  {journal} {\bibinfo
  {journal} {Physical Review Letters}\ }\textbf {\bibinfo {volume} {106}},\
  \bibinfo {pages} {123601} (\bibinfo {year} {2011})}\BibitemShut {NoStop}%
\bibitem [{\citenamefont {Schultze}\ \emph {et~al.}(2014)\citenamefont
  {Schultze}, \citenamefont {Ramasesha}, \citenamefont {Pemmaraju},
  \citenamefont {Sato}, \citenamefont {Whitmore}, \citenamefont {Gandman},
  \citenamefont {Prell}, \citenamefont {Borja}, \citenamefont {Prendergast},
  \citenamefont {Yabana}, \citenamefont {Neumark},\ and\ \citenamefont
  {Leone}}]{schultze14}%
  \BibitemOpen
  \bibfield  {author} {\bibinfo {author} {\bibfnamefont {M.}~\bibnamefont
  {Schultze}}, \bibinfo {author} {\bibfnamefont {K.}~\bibnamefont {Ramasesha}},
  \bibinfo {author} {\bibfnamefont {C.~D.}\ \bibnamefont {Pemmaraju}}, \bibinfo
  {author} {\bibfnamefont {S.~A.}\ \bibnamefont {Sato}}, \bibinfo {author}
  {\bibfnamefont {D.}~\bibnamefont {Whitmore}}, \bibinfo {author}
  {\bibfnamefont {A.}~\bibnamefont {Gandman}}, \bibinfo {author} {\bibfnamefont
  {J.~S.}\ \bibnamefont {Prell}}, \bibinfo {author} {\bibfnamefont {L.~J.}\
  \bibnamefont {Borja}}, \bibinfo {author} {\bibfnamefont {D.}~\bibnamefont
  {Prendergast}}, \bibinfo {author} {\bibfnamefont {K.}~\bibnamefont {Yabana}},
  \bibinfo {author} {\bibfnamefont {D.~M.}\ \bibnamefont {Neumark}}, \ and\
  \bibinfo {author} {\bibfnamefont {S.~R.}\ \bibnamefont {Leone}},\ }\href
  {\doibase 10.1126/science.1260311} {\bibfield  {journal} {\bibinfo  {journal}
  {Science}\ }\textbf {\bibinfo {volume} {346}},\ \bibinfo {pages} {1348}
  (\bibinfo {year} {2014})}\BibitemShut {NoStop}%
\bibitem [{\citenamefont {Lucchini}\ \emph {et~al.}(2016)\citenamefont
  {Lucchini}, \citenamefont {Sato}, \citenamefont {Ludwig}, \citenamefont
  {Herrmann}, \citenamefont {Volkov}, \citenamefont {Kasmi}, \citenamefont
  {Shinohara}, \citenamefont {Yabana}, \citenamefont {Gallmann},\ and\
  \citenamefont {Keller}}]{lucchini16}%
  \BibitemOpen
  \bibfield  {author} {\bibinfo {author} {\bibfnamefont {M.}~\bibnamefont
  {Lucchini}}, \bibinfo {author} {\bibfnamefont {S.~A.}\ \bibnamefont {Sato}},
  \bibinfo {author} {\bibfnamefont {A.}~\bibnamefont {Ludwig}}, \bibinfo
  {author} {\bibfnamefont {J.}~\bibnamefont {Herrmann}}, \bibinfo {author}
  {\bibfnamefont {M.}~\bibnamefont {Volkov}}, \bibinfo {author} {\bibfnamefont
  {L.}~\bibnamefont {Kasmi}}, \bibinfo {author} {\bibfnamefont
  {Y.}~\bibnamefont {Shinohara}}, \bibinfo {author} {\bibfnamefont
  {K.}~\bibnamefont {Yabana}}, \bibinfo {author} {\bibfnamefont
  {L.}~\bibnamefont {Gallmann}}, \ and\ \bibinfo {author} {\bibfnamefont
  {U.}~\bibnamefont {Keller}},\ }\href {\doibase 10.1126/science.aag1268}
  {\bibfield  {journal} {\bibinfo  {journal} {Science}\ }\textbf {\bibinfo
  {volume} {353}},\ \bibinfo {pages} {916} (\bibinfo {year}
  {2016})}\BibitemShut {NoStop}%
\bibitem [{\citenamefont {Chen}\ \emph {et~al.}(2012)\citenamefont {Chen},
  \citenamefont {Bell}, \citenamefont {Beck}, \citenamefont {Mashiko},
  \citenamefont {Wu}, \citenamefont {Pfeiffer}, \citenamefont {Gaarde},
  \citenamefont {Neumark}, \citenamefont {Leone},\ and\ \citenamefont
  {Schafer}}]{chen12}%
  \BibitemOpen
  \bibfield  {author} {\bibinfo {author} {\bibfnamefont {S.}~\bibnamefont
  {Chen}}, \bibinfo {author} {\bibfnamefont {M.~J.}\ \bibnamefont {Bell}},
  \bibinfo {author} {\bibfnamefont {A.~R.}\ \bibnamefont {Beck}}, \bibinfo
  {author} {\bibfnamefont {H.}~\bibnamefont {Mashiko}}, \bibinfo {author}
  {\bibfnamefont {M.}~\bibnamefont {Wu}}, \bibinfo {author} {\bibfnamefont
  {A.~N.}\ \bibnamefont {Pfeiffer}}, \bibinfo {author} {\bibfnamefont {M.~B.}\
  \bibnamefont {Gaarde}}, \bibinfo {author} {\bibfnamefont {D.~M.}\
  \bibnamefont {Neumark}}, \bibinfo {author} {\bibfnamefont {S.~R.}\
  \bibnamefont {Leone}}, \ and\ \bibinfo {author} {\bibfnamefont {K.~J.}\
  \bibnamefont {Schafer}},\ }\href {\doibase 10.1103/PhysRevA.86.063408}
  {\bibfield  {journal} {\bibinfo  {journal} {Physical Review A}\ }\textbf
  {\bibinfo {volume} {86}},\ \bibinfo {pages} {063408} (\bibinfo {year}
  {2012})}\BibitemShut {NoStop}%
\bibitem [{\citenamefont {Reduzzi}\ \emph {et~al.}(2015)\citenamefont
  {Reduzzi}, \citenamefont {Hummert}, \citenamefont {Dubrouil}, \citenamefont
  {Calegari}, \citenamefont {Nisoli}, \citenamefont {Frassetto}, \citenamefont
  {Poletto}, \citenamefont {Chen}, \citenamefont {Wu}, \citenamefont {Gaarde},
  \citenamefont {Schafer},\ and\ \citenamefont {Sansone}}]{reduzzi15}%
  \BibitemOpen
  \bibfield  {author} {\bibinfo {author} {\bibfnamefont {M.}~\bibnamefont
  {Reduzzi}}, \bibinfo {author} {\bibfnamefont {J.}~\bibnamefont {Hummert}},
  \bibinfo {author} {\bibfnamefont {A.}~\bibnamefont {Dubrouil}}, \bibinfo
  {author} {\bibfnamefont {F.}~\bibnamefont {Calegari}}, \bibinfo {author}
  {\bibfnamefont {M.}~\bibnamefont {Nisoli}}, \bibinfo {author} {\bibfnamefont
  {F.}~\bibnamefont {Frassetto}}, \bibinfo {author} {\bibfnamefont
  {L.}~\bibnamefont {Poletto}}, \bibinfo {author} {\bibfnamefont
  {S.}~\bibnamefont {Chen}}, \bibinfo {author} {\bibfnamefont {M.}~\bibnamefont
  {Wu}}, \bibinfo {author} {\bibfnamefont {M.~B.}\ \bibnamefont {Gaarde}},
  \bibinfo {author} {\bibfnamefont {K.}~\bibnamefont {Schafer}}, \ and\
  \bibinfo {author} {\bibfnamefont {G.}~\bibnamefont {Sansone}},\ }\href
  {\doibase 10.1103/PhysRevA.92.033408} {\bibfield  {journal} {\bibinfo
  {journal} {Physical Review A}\ }\textbf {\bibinfo {volume} {92}},\ \bibinfo
  {pages} {033408} (\bibinfo {year} {2015})}\BibitemShut {NoStop}%
\bibitem [{\citenamefont {Chini}\ \emph {et~al.}(2012)\citenamefont {Chini},
  \citenamefont {Zhao}, \citenamefont {Wang}, \citenamefont {Cheng},
  \citenamefont {Hu},\ and\ \citenamefont {Chang}}]{chini12}%
  \BibitemOpen
  \bibfield  {author} {\bibinfo {author} {\bibfnamefont {M.}~\bibnamefont
  {Chini}}, \bibinfo {author} {\bibfnamefont {B.}~\bibnamefont {Zhao}},
  \bibinfo {author} {\bibfnamefont {H.}~\bibnamefont {Wang}}, \bibinfo {author}
  {\bibfnamefont {Y.}~\bibnamefont {Cheng}}, \bibinfo {author} {\bibfnamefont
  {S.~X.}\ \bibnamefont {Hu}}, \ and\ \bibinfo {author} {\bibfnamefont
  {Z.}~\bibnamefont {Chang}},\ }\href {\doibase 10.1103/PhysRevLett.109.073601}
  {\bibfield  {journal} {\bibinfo  {journal} {Physical Review Letters}\
  }\textbf {\bibinfo {volume} {109}},\ \bibinfo {pages} {073601} (\bibinfo
  {year} {2012})}\BibitemShut {NoStop}%
\bibitem [{\citenamefont {Beck}\ \emph {et~al.}(2014)\citenamefont {Beck},
  \citenamefont {Bernhardt}, \citenamefont {Warrick}, \citenamefont {Wu},
  \citenamefont {Chen}, \citenamefont {Gaarde}, \citenamefont {Schafer},
  \citenamefont {Neumark},\ and\ \citenamefont {Leone}}]{beck14}%
  \BibitemOpen
  \bibfield  {author} {\bibinfo {author} {\bibfnamefont {A.~R.}\ \bibnamefont
  {Beck}}, \bibinfo {author} {\bibfnamefont {B.}~\bibnamefont {Bernhardt}},
  \bibinfo {author} {\bibfnamefont {E.~R.}\ \bibnamefont {Warrick}}, \bibinfo
  {author} {\bibfnamefont {M.}~\bibnamefont {Wu}}, \bibinfo {author}
  {\bibfnamefont {S.}~\bibnamefont {Chen}}, \bibinfo {author} {\bibfnamefont
  {M.~B.}\ \bibnamefont {Gaarde}}, \bibinfo {author} {\bibfnamefont {K.~J.}\
  \bibnamefont {Schafer}}, \bibinfo {author} {\bibfnamefont {D.~M.}\
  \bibnamefont {Neumark}}, \ and\ \bibinfo {author} {\bibfnamefont {S.~R.}\
  \bibnamefont {Leone}},\ }\href {\doibase 10.1088/1367-2630/16/11/113016}
  {\bibfield  {journal} {\bibinfo  {journal} {New Journal of Physics}\ }\textbf
  {\bibinfo {volume} {16}},\ \bibinfo {pages} {113016} (\bibinfo {year}
  {2014})}\BibitemShut {NoStop}%
\bibitem [{\citenamefont {Cao}\ \emph {et~al.}(2016{\natexlab{a}})\citenamefont
  {Cao}, \citenamefont {Warrick}, \citenamefont {Neumark},\ and\ \citenamefont
  {Leone}}]{cao16}%
  \BibitemOpen
  \bibfield  {author} {\bibinfo {author} {\bibfnamefont {W.}~\bibnamefont
  {Cao}}, \bibinfo {author} {\bibfnamefont {E.~R.}\ \bibnamefont {Warrick}},
  \bibinfo {author} {\bibfnamefont {D.~M.}\ \bibnamefont {Neumark}}, \ and\
  \bibinfo {author} {\bibfnamefont {S.~R.}\ \bibnamefont {Leone}},\ }\href
  {\doibase 10.1088/1367-2630/18/1/013041} {\bibfield  {journal} {\bibinfo
  {journal} {New Journal of Physics}\ }\textbf {\bibinfo {volume} {18}},\
  \bibinfo {pages} {013041} (\bibinfo {year} {2016}{\natexlab{a}})}\BibitemShut
  {NoStop}%
\bibitem [{\citenamefont {Cheng}\ \emph {et~al.}(2016)\citenamefont {Cheng},
  \citenamefont {Chini}, \citenamefont {Wang}, \citenamefont
  {Gonz\'alez-Castrillo}, \citenamefont {Palacios}, \citenamefont {Argenti},
  \citenamefont {Mart\'{\i}n},\ and\ \citenamefont {Chang}}]{cheng16}%
  \BibitemOpen
  \bibfield  {author} {\bibinfo {author} {\bibfnamefont {Y.}~\bibnamefont
  {Cheng}}, \bibinfo {author} {\bibfnamefont {M.}~\bibnamefont {Chini}},
  \bibinfo {author} {\bibfnamefont {X.}~\bibnamefont {Wang}}, \bibinfo {author}
  {\bibfnamefont {A.}~\bibnamefont {Gonz\'alez-Castrillo}}, \bibinfo {author}
  {\bibfnamefont {A.}~\bibnamefont {Palacios}}, \bibinfo {author}
  {\bibfnamefont {L.}~\bibnamefont {Argenti}}, \bibinfo {author} {\bibfnamefont
  {F.}~\bibnamefont {Mart\'{\i}n}}, \ and\ \bibinfo {author} {\bibfnamefont
  {Z.}~\bibnamefont {Chang}},\ }\href {\doibase 10.1103/PhysRevA.94.023403}
  {\bibfield  {journal} {\bibinfo  {journal} {Phys. Rev. A}\ }\textbf {\bibinfo
  {volume} {94}},\ \bibinfo {pages} {023403} (\bibinfo {year}
  {2016})}\BibitemShut {NoStop}%
\bibitem [{\citenamefont {Warrick}\ \emph {et~al.}(2016)\citenamefont
  {Warrick}, \citenamefont {Cao}, \citenamefont {Neumark},\ and\ \citenamefont
  {Leone}}]{warrick16}%
  \BibitemOpen
  \bibfield  {author} {\bibinfo {author} {\bibfnamefont {E.~R.}\ \bibnamefont
  {Warrick}}, \bibinfo {author} {\bibfnamefont {W.}~\bibnamefont {Cao}},
  \bibinfo {author} {\bibfnamefont {D.~M.}\ \bibnamefont {Neumark}}, \ and\
  \bibinfo {author} {\bibfnamefont {S.~R.}\ \bibnamefont {Leone}},\ }\href
  {\doibase 10.1021/acs.jpca.5b11570} {\bibfield  {journal} {\bibinfo
  {journal} {The Journal of Physical Chemistry A}\ }\textbf {\bibinfo {volume}
  {120}},\ \bibinfo {pages} {3165} (\bibinfo {year} {2016})}\BibitemShut
  {NoStop}%
\bibitem [{\citenamefont {Ott}\ \emph {et~al.}(2013)\citenamefont {Ott},
  \citenamefont {Kaldun}, \citenamefont {Raith}, \citenamefont {Meyer},
  \citenamefont {Laux}, \citenamefont {Evers}, \citenamefont {Keitel},
  \citenamefont {Greene},\ and\ \citenamefont {Pfeifer}}]{ott13}%
  \BibitemOpen
  \bibfield  {author} {\bibinfo {author} {\bibfnamefont {C.}~\bibnamefont
  {Ott}}, \bibinfo {author} {\bibfnamefont {A.}~\bibnamefont {Kaldun}},
  \bibinfo {author} {\bibfnamefont {P.}~\bibnamefont {Raith}}, \bibinfo
  {author} {\bibfnamefont {K.}~\bibnamefont {Meyer}}, \bibinfo {author}
  {\bibfnamefont {M.}~\bibnamefont {Laux}}, \bibinfo {author} {\bibfnamefont
  {J.}~\bibnamefont {Evers}}, \bibinfo {author} {\bibfnamefont {C.~H.}\
  \bibnamefont {Keitel}}, \bibinfo {author} {\bibfnamefont {C.~H.}\
  \bibnamefont {Greene}}, \ and\ \bibinfo {author} {\bibfnamefont
  {T.}~\bibnamefont {Pfeifer}},\ }\href {\doibase 10.1126/science.1234407}
  {\bibfield  {journal} {\bibinfo  {journal} {Science}\ }\textbf {\bibinfo
  {volume} {340}},\ \bibinfo {pages} {716} (\bibinfo {year}
  {2013})}\BibitemShut {NoStop}%
\bibitem [{\citenamefont {Bengtsson}\ \emph {et~al.}(2016)\citenamefont
  {Bengtsson}, \citenamefont {Larsen}, \citenamefont {Kroon}, \citenamefont
  {Camp}, \citenamefont {Miranda}, \citenamefont {Arnold}, \citenamefont
  {L'Huillier}, \citenamefont {Schafer}, \citenamefont {Gaarde}, \citenamefont
  {Rippe},\ and\ \citenamefont {Mauritsson}}]{bengtsson16}%
  \BibitemOpen
  \bibfield  {author} {\bibinfo {author} {\bibfnamefont {S.}~\bibnamefont
  {Bengtsson}}, \bibinfo {author} {\bibfnamefont {E.~W.}\ \bibnamefont
  {Larsen}}, \bibinfo {author} {\bibfnamefont {D.}~\bibnamefont {Kroon}},
  \bibinfo {author} {\bibfnamefont {S.}~\bibnamefont {Camp}}, \bibinfo {author}
  {\bibfnamefont {M.}~\bibnamefont {Miranda}}, \bibinfo {author} {\bibfnamefont
  {C.~L.}\ \bibnamefont {Arnold}}, \bibinfo {author} {\bibfnamefont
  {A.}~\bibnamefont {L'Huillier}}, \bibinfo {author} {\bibfnamefont {K.~J.}\
  \bibnamefont {Schafer}}, \bibinfo {author} {\bibfnamefont {M.~B.}\
  \bibnamefont {Gaarde}}, \bibinfo {author} {\bibfnamefont {L.}~\bibnamefont
  {Rippe}}, \ and\ \bibinfo {author} {\bibfnamefont {J.}~\bibnamefont
  {Mauritsson}},\ }\href {http://arxiv.org/abs/1611.04836} {\bibfield
  {journal} {\bibinfo  {journal} {arXiv:1611.04836 [physics]}\ } (\bibinfo
  {year} {2016})},\ \bibinfo {note} {arXiv: 1611.04836}\BibitemShut {NoStop}%
\bibitem [{\citenamefont {Ott}\ \emph {et~al.}(2014)\citenamefont {Ott},
  \citenamefont {Kaldun}, \citenamefont {Argenti}, \citenamefont {Raith},
  \citenamefont {Meyer}, \citenamefont {Laux}, \citenamefont {Zhang},
  \citenamefont {Blättermann}, \citenamefont {Hagstotz}, \citenamefont {Ding},
  \citenamefont {Heck}, \citenamefont {Madroñero}, \citenamefont {Martín},\
  and\ \citenamefont {Pfeifer}}]{ott14}%
  \BibitemOpen
  \bibfield  {author} {\bibinfo {author} {\bibfnamefont {C.}~\bibnamefont
  {Ott}}, \bibinfo {author} {\bibfnamefont {A.}~\bibnamefont {Kaldun}},
  \bibinfo {author} {\bibfnamefont {L.}~\bibnamefont {Argenti}}, \bibinfo
  {author} {\bibfnamefont {P.}~\bibnamefont {Raith}}, \bibinfo {author}
  {\bibfnamefont {K.}~\bibnamefont {Meyer}}, \bibinfo {author} {\bibfnamefont
  {M.}~\bibnamefont {Laux}}, \bibinfo {author} {\bibfnamefont {Y.}~\bibnamefont
  {Zhang}}, \bibinfo {author} {\bibfnamefont {A.}~\bibnamefont {Blättermann}},
  \bibinfo {author} {\bibfnamefont {S.}~\bibnamefont {Hagstotz}}, \bibinfo
  {author} {\bibfnamefont {T.}~\bibnamefont {Ding}}, \bibinfo {author}
  {\bibfnamefont {R.}~\bibnamefont {Heck}}, \bibinfo {author} {\bibfnamefont
  {J.}~\bibnamefont {Madroñero}}, \bibinfo {author} {\bibfnamefont
  {F.}~\bibnamefont {Martín}}, \ and\ \bibinfo {author} {\bibfnamefont
  {T.}~\bibnamefont {Pfeifer}},\ }\href {\doibase 10.1038/nature14026}
  {\bibfield  {journal} {\bibinfo  {journal} {Nature}\ }\textbf {\bibinfo
  {volume} {516}},\ \bibinfo {pages} {374} (\bibinfo {year}
  {2014})}\BibitemShut {NoStop}%
\bibitem [{\citenamefont {Liao}\ \emph {et~al.}(2015)\citenamefont {Liao},
  \citenamefont {Sandhu}, \citenamefont {Camp}, \citenamefont {Schafer},\ and\
  \citenamefont {Gaarde}}]{liao15}%
  \BibitemOpen
  \bibfield  {author} {\bibinfo {author} {\bibfnamefont {C.-T.}\ \bibnamefont
  {Liao}}, \bibinfo {author} {\bibfnamefont {A.}~\bibnamefont {Sandhu}},
  \bibinfo {author} {\bibfnamefont {S.}~\bibnamefont {Camp}}, \bibinfo {author}
  {\bibfnamefont {K.~J.}\ \bibnamefont {Schafer}}, \ and\ \bibinfo {author}
  {\bibfnamefont {M.~B.}\ \bibnamefont {Gaarde}},\ }\href {\doibase
  10.1103/PhysRevLett.114.143002} {\bibfield  {journal} {\bibinfo  {journal}
  {Physical Review Letters}\ }\textbf {\bibinfo {volume} {114}},\ \bibinfo
  {pages} {143002} (\bibinfo {year} {2015})}\BibitemShut {NoStop}%
\bibitem [{\citenamefont {Beaulieu}\ \emph {et~al.}(2016)\citenamefont
  {Beaulieu}, \citenamefont {Camp}, \citenamefont {Descamps}, \citenamefont
  {Comby}, \citenamefont {Wanie}, \citenamefont {Petit}, \citenamefont
  {Légaré}, \citenamefont {Schafer}, \citenamefont {Gaarde}, \citenamefont
  {Catoire},\ and\ \citenamefont {Mairesse}}]{beaulieu16}%
  \BibitemOpen
  \bibfield  {author} {\bibinfo {author} {\bibfnamefont {S.}~\bibnamefont
  {Beaulieu}}, \bibinfo {author} {\bibfnamefont {S.}~\bibnamefont {Camp}},
  \bibinfo {author} {\bibfnamefont {D.}~\bibnamefont {Descamps}}, \bibinfo
  {author} {\bibfnamefont {A.}~\bibnamefont {Comby}}, \bibinfo {author}
  {\bibfnamefont {V.}~\bibnamefont {Wanie}}, \bibinfo {author} {\bibfnamefont
  {S.}~\bibnamefont {Petit}}, \bibinfo {author} {\bibfnamefont
  {F.}~\bibnamefont {L\'egar\'e}}, \bibinfo {author} {\bibfnamefont
  {K.}~\bibnamefont {Schafer}}, \bibinfo {author} {\bibfnamefont
  {M.}~\bibnamefont {Gaarde}}, \bibinfo {author} {\bibfnamefont
  {F.}~\bibnamefont {Catoire}}, \ and\ \bibinfo {author} {\bibfnamefont
  {Y.}~\bibnamefont {Mairesse}},\ }\href {\doibase
  10.1103/PhysRevLett.117.203001} {\bibfield  {journal} {\bibinfo  {journal}
  {Physical Review Letters}\ }\textbf {\bibinfo {volume} {117}},\ \bibinfo
  {pages} {203001} (\bibinfo {year} {2016})}\BibitemShut {NoStop}%
\bibitem [{\citenamefont {Bellini}\ \emph {et~al.}(1998)\citenamefont
  {Bellini}, \citenamefont {Lyngå}, \citenamefont {Tozzi}, \citenamefont
  {Gaarde}, \citenamefont {Hänsch}, \citenamefont {L'Huillier},\ and\
  \citenamefont {Wahlström}}]{bellini98}%
  \BibitemOpen
  \bibfield  {author} {\bibinfo {author} {\bibfnamefont {M.}~\bibnamefont
  {Bellini}}, \bibinfo {author} {\bibfnamefont {C.}~\bibnamefont {Lyngå}},
  \bibinfo {author} {\bibfnamefont {A.}~\bibnamefont {Tozzi}}, \bibinfo
  {author} {\bibfnamefont {M.~B.}\ \bibnamefont {Gaarde}}, \bibinfo {author}
  {\bibfnamefont {T.~W.}\ \bibnamefont {Hänsch}}, \bibinfo {author}
  {\bibfnamefont {A.}~\bibnamefont {L'Huillier}}, \ and\ \bibinfo {author}
  {\bibfnamefont {C.-G.}\ \bibnamefont {Wahlström}},\ }\href {\doibase
  10.1103/PhysRevLett.81.297} {\bibfield  {journal} {\bibinfo  {journal}
  {Physical Review Letters}\ }\textbf {\bibinfo {volume} {81}},\ \bibinfo
  {pages} {297} (\bibinfo {year} {1998})}\BibitemShut {NoStop}%
\bibitem [{\citenamefont {Yoshino}(1970)}]{yoshino70}%
  \BibitemOpen
  \bibfield  {author} {\bibinfo {author} {\bibfnamefont {K.}~\bibnamefont
  {Yoshino}},\ }\href {\doibase 10.1364/JOSA.60.001220} {\bibfield  {journal}
  {\bibinfo  {journal} {JOSA}\ }\textbf {\bibinfo {volume} {60}},\ \bibinfo
  {pages} {1220} (\bibinfo {year} {1970})}\BibitemShut {NoStop}%
\bibitem [{\citenamefont {Chen}\ \emph {et~al.}(2013)\citenamefont {Chen},
  \citenamefont {Wu}, \citenamefont {Gaarde},\ and\ \citenamefont
  {Schafer}}]{chen13}%
  \BibitemOpen
  \bibfield  {author} {\bibinfo {author} {\bibfnamefont {S.}~\bibnamefont
  {Chen}}, \bibinfo {author} {\bibfnamefont {M.}~\bibnamefont {Wu}}, \bibinfo
  {author} {\bibfnamefont {M.~B.}\ \bibnamefont {Gaarde}}, \ and\ \bibinfo
  {author} {\bibfnamefont {K.~J.}\ \bibnamefont {Schafer}},\ }\href {\doibase
  10.1103/PhysRevA.87.033408} {\bibfield  {journal} {\bibinfo  {journal}
  {Physical Review A}\ }\textbf {\bibinfo {volume} {87}},\ \bibinfo {pages}
  {033408} (\bibinfo {year} {2013})}\BibitemShut {NoStop}%
\bibitem [{\citenamefont {Cao}\ \emph {et~al.}(2016{\natexlab{b}})\citenamefont
  {Cao}, \citenamefont {Warrick}, \citenamefont {Fidler}, \citenamefont
  {Leone},\ and\ \citenamefont {Neumark}}]{cao16-1}%
  \BibitemOpen
  \bibfield  {author} {\bibinfo {author} {\bibfnamefont {W.}~\bibnamefont
  {Cao}}, \bibinfo {author} {\bibfnamefont {E.~R.}\ \bibnamefont {Warrick}},
  \bibinfo {author} {\bibfnamefont {A.}~\bibnamefont {Fidler}}, \bibinfo
  {author} {\bibfnamefont {S.~R.}\ \bibnamefont {Leone}}, \ and\ \bibinfo
  {author} {\bibfnamefont {D.~M.}\ \bibnamefont {Neumark}},\ }\href {\doibase
  10.1103/PhysRevA.94.021802} {\bibfield  {journal} {\bibinfo  {journal}
  {Physical Review A}\ }\textbf {\bibinfo {volume} {94}},\ \bibinfo {pages}
  {021802} (\bibinfo {year} {2016}{\natexlab{b}})}\BibitemShut {NoStop}%
\bibitem [{\citenamefont {Cao}\ \emph {et~al.}(2016{\natexlab{c}})\citenamefont
  {Cao}, \citenamefont {Warrick}, \citenamefont {Fidler}, \citenamefont
  {Neumark},\ and\ \citenamefont {Leone}}]{cao16-2}%
  \BibitemOpen
  \bibfield  {author} {\bibinfo {author} {\bibfnamefont {W.}~\bibnamefont
  {Cao}}, \bibinfo {author} {\bibfnamefont {E.~R.}\ \bibnamefont {Warrick}},
  \bibinfo {author} {\bibfnamefont {A.}~\bibnamefont {Fidler}}, \bibinfo
  {author} {\bibfnamefont {D.~M.}\ \bibnamefont {Neumark}}, \ and\ \bibinfo
  {author} {\bibfnamefont {S.~R.}\ \bibnamefont {Leone}},\ }\href {\doibase
  10.1103/PhysRevA.94.053846} {\bibfield  {journal} {\bibinfo  {journal}
  {Physical Review A}\ }\textbf {\bibinfo {volume} {94}},\ \bibinfo {pages}
  {053846} (\bibinfo {year} {2016}{\natexlab{c}})}\BibitemShut {NoStop}%
\bibitem [{\citenamefont {Camper}\ \emph {et~al.}(2015)\citenamefont {Camper},
  \citenamefont {Ferr\'e}, \citenamefont {Lin}, \citenamefont {Skantzakis},
  \citenamefont {Staedter}, \citenamefont {English}, \citenamefont
  {Manschwetus}, \citenamefont {Burgy}, \citenamefont {Petit}, \citenamefont
  {Descamps}, \citenamefont {Auguste}, \citenamefont {Gobert}, \citenamefont
  {Carr\'e}, \citenamefont {Sali\'eres}, \citenamefont {Mairesse},\ and\
  \citenamefont {Ruchon}}]{camper15}%
  \BibitemOpen
  \bibfield  {author} {\bibinfo {author} {\bibfnamefont {A.}~\bibnamefont
  {Camper}}, \bibinfo {author} {\bibfnamefont {A.}~\bibnamefont {Ferré}},
  \bibinfo {author} {\bibfnamefont {N.}~\bibnamefont {Lin}}, \bibinfo {author}
  {\bibfnamefont {E.}~\bibnamefont {Skantzakis}}, \bibinfo {author}
  {\bibfnamefont {D.}~\bibnamefont {Staedter}}, \bibinfo {author}
  {\bibfnamefont {E.}~\bibnamefont {English}}, \bibinfo {author} {\bibfnamefont
  {B.}~\bibnamefont {Manschwetus}}, \bibinfo {author} {\bibfnamefont
  {F.}~\bibnamefont {Burgy}}, \bibinfo {author} {\bibfnamefont
  {S.}~\bibnamefont {Petit}}, \bibinfo {author} {\bibfnamefont
  {D.}~\bibnamefont {Descamps}}, \bibinfo {author} {\bibfnamefont
  {T.}~\bibnamefont {Auguste}}, \bibinfo {author} {\bibfnamefont
  {O.}~\bibnamefont {Gobert}}, \bibinfo {author} {\bibfnamefont
  {B.}~\bibnamefont {Carré}}, \bibinfo {author} {\bibfnamefont
  {P.}~\bibnamefont {Salières}}, \bibinfo {author} {\bibfnamefont
  {Y.}~\bibnamefont {Mairesse}}, \ and\ \bibinfo {author} {\bibfnamefont
  {T.}~\bibnamefont {Ruchon}},\ }\href {\doibase 10.3390/photonics2010184}
  {\bibfield  {journal} {\bibinfo  {journal} {Photonics}\ }\textbf {\bibinfo
  {volume} {2}},\ \bibinfo {pages} {184} (\bibinfo {year} {2015})}\BibitemShut
  {NoStop}%
\bibitem [{\citenamefont {Young}\ \emph {et~al.}(2006)\citenamefont {Young},
  \citenamefont {Arms}, \citenamefont {Dufresne}, \citenamefont {Dunford},
  \citenamefont {Ederer}, \citenamefont {Höhr}, \citenamefont {Kanter},
  \citenamefont {Krässig}, \citenamefont {Landahl}, \citenamefont {Peterson},
  \citenamefont {Rudati}, \citenamefont {Santra},\ and\ \citenamefont
  {Southworth}}]{young06}%
  \BibitemOpen
  \bibfield  {author} {\bibinfo {author} {\bibfnamefont {L.}~\bibnamefont
  {Young}}, \bibinfo {author} {\bibfnamefont {D.~A.}\ \bibnamefont {Arms}},
  \bibinfo {author} {\bibfnamefont {E.~M.}\ \bibnamefont {Dufresne}}, \bibinfo
  {author} {\bibfnamefont {R.~W.}\ \bibnamefont {Dunford}}, \bibinfo {author}
  {\bibfnamefont {D.~L.}\ \bibnamefont {Ederer}}, \bibinfo {author}
  {\bibfnamefont {C.}~\bibnamefont {Höhr}}, \bibinfo {author} {\bibfnamefont
  {E.~P.}\ \bibnamefont {Kanter}}, \bibinfo {author} {\bibfnamefont
  {B.}~\bibnamefont {Krässig}}, \bibinfo {author} {\bibfnamefont {E.~C.}\
  \bibnamefont {Landahl}}, \bibinfo {author} {\bibfnamefont {E.~R.}\
  \bibnamefont {Peterson}}, \bibinfo {author} {\bibfnamefont {J.}~\bibnamefont
  {Rudati}}, \bibinfo {author} {\bibfnamefont {R.}~\bibnamefont {Santra}}, \
  and\ \bibinfo {author} {\bibfnamefont {S.~H.}\ \bibnamefont {Southworth}},\
  }\href {\doibase 10.1103/PhysRevLett.97.083601} {\bibfield  {journal}
  {\bibinfo  {journal} {Physical Review Letters}\ }\textbf {\bibinfo {volume}
  {97}},\ \bibinfo {pages} {083601} (\bibinfo {year} {2006})}\BibitemShut
  {NoStop}%
\bibitem [{\citenamefont {Pavicic}\ \emph {et~al.}(2007)\citenamefont
  {Pavicic}, \citenamefont {Lee}, \citenamefont {Rayner}, \citenamefont
  {Corkum},\ and\ \citenamefont {Villeneuve}}]{pavicic07}%
  \BibitemOpen
  \bibfield  {author} {\bibinfo {author} {\bibfnamefont {D.}~\bibnamefont
  {Pavičić}}, \bibinfo {author} {\bibfnamefont {K.~F.}\ \bibnamefont {Lee}},
  \bibinfo {author} {\bibfnamefont {D.~M.}\ \bibnamefont {Rayner}}, \bibinfo
  {author} {\bibfnamefont {P.~B.}\ \bibnamefont {Corkum}}, \ and\ \bibinfo
  {author} {\bibfnamefont {D.~M.}\ \bibnamefont {Villeneuve}},\ }\href
  {\doibase 10.1103/PhysRevLett.98.243001} {\bibfield  {journal} {\bibinfo
  {journal} {Physical Review Letters}\ }\textbf {\bibinfo {volume} {98}},\
  \bibinfo {pages} {243001} (\bibinfo {year} {2007})}\BibitemShut {NoStop}%
\bibitem [{\citenamefont {Shafir}\ \emph {et~al.}(2009)\citenamefont {Shafir},
  \citenamefont {Mairesse}, \citenamefont {Villeneuve}, \citenamefont
  {Corkum},\ and\ \citenamefont {Dudovich}}]{shafir09}%
  \BibitemOpen
  \bibfield  {author} {\bibinfo {author} {\bibfnamefont {D.}~\bibnamefont
  {Shafir}}, \bibinfo {author} {\bibfnamefont {Y.}~\bibnamefont {Mairesse}},
  \bibinfo {author} {\bibfnamefont {D.~M.}\ \bibnamefont {Villeneuve}},
  \bibinfo {author} {\bibfnamefont {P.~B.}\ \bibnamefont {Corkum}}, \ and\
  \bibinfo {author} {\bibfnamefont {N.}~\bibnamefont {Dudovich}},\ }\href
  {\doibase 10.1038/nphys1251} {\bibfield  {journal} {\bibinfo  {journal}
  {Nature Physics}\ }\textbf {\bibinfo {volume} {5}},\ \bibinfo {pages} {412}
  (\bibinfo {year} {2009})}\BibitemShut {NoStop}%
\bibitem [{\citenamefont {Potvliege}\ and\ \citenamefont
  {Vučić}(2006)}]{potvliege06}%
  \BibitemOpen
  \bibfield  {author} {\bibinfo {author} {\bibfnamefont {R.~M.}\ \bibnamefont
  {Potvliege}}\ and\ \bibinfo {author} {\bibfnamefont {S.}~\bibnamefont
  {Vučić}},\ }\href {\doibase 10.1103/PhysRevA.74.023412} {\bibfield
  {journal} {\bibinfo  {journal} {Physical Review A}\ }\textbf {\bibinfo
  {volume} {74}},\ \bibinfo {pages} {023412} (\bibinfo {year}
  {2006})}\BibitemShut {NoStop}%
\bibitem [{\citenamefont {L\'opez-Martens}\ \emph {et~al.}(2000)\citenamefont
  {López-Martens}, \citenamefont {Schmidt},\ and\ \citenamefont
  {Roberts}}]{lopez-martens00}%
  \BibitemOpen
  \bibfield  {author} {\bibinfo {author} {\bibfnamefont {R.~B.}\ \bibnamefont
  {L\'opez-Martens}}, \bibinfo {author} {\bibfnamefont {T.~W.}\ \bibnamefont
  {Schmidt}}, \ and\ \bibinfo {author} {\bibfnamefont {G.}~\bibnamefont
  {Roberts}},\ }\href {\doibase 10.1103/PhysRevA.62.013414} {\bibfield
  {journal} {\bibinfo  {journal} {Physical Review A}\ }\textbf {\bibinfo
  {volume} {62}},\ \bibinfo {pages} {013414} (\bibinfo {year}
  {2000})}\BibitemShut {NoStop}%
\end{thebibliography}%

\clearpage

\textbf{Supplementary Materials} \\
\textbf{High-resolution 2D spectra} 

The XUV spectrometer used in the experiment has a resolution of $\sim$ 60 $meV$ around 15 eV. Because different spin-orbit Rydberg states of argon are separated by less than 10-20 meV, it is impossible to unambiguously assign them from the measurement of the XFID spectrum. By using interference between $\Lambda$- or ladder-type transitions, we performed high-resolution Fourier-transform spectroscopy to obtain 2D spectra. The measurements shown in the main text (Fig. 2 from main text) were done scanning the delay between 0.1 and 0.5 ps, i.e. over a range $\Delta\tau=0.4$ ps. The spectral resolution was thus $\delta\nu=1/\Delta\tau\approx 10$ meV. We performed additional measurements with a larger scanning range, from 0.1 to 2.4 ps. The resulting 2D spectrum shown in Fig. 1 enables determining the frequencies of the oscillations with a resolution $\Delta E$=1.72 meV. This resolution could be further improved by increasing the pump-probe delay scanning range. \\

\hspace{1cm}
\begin{figure}[H]
\begin{center}
\includegraphics[width=0.40\textwidth]{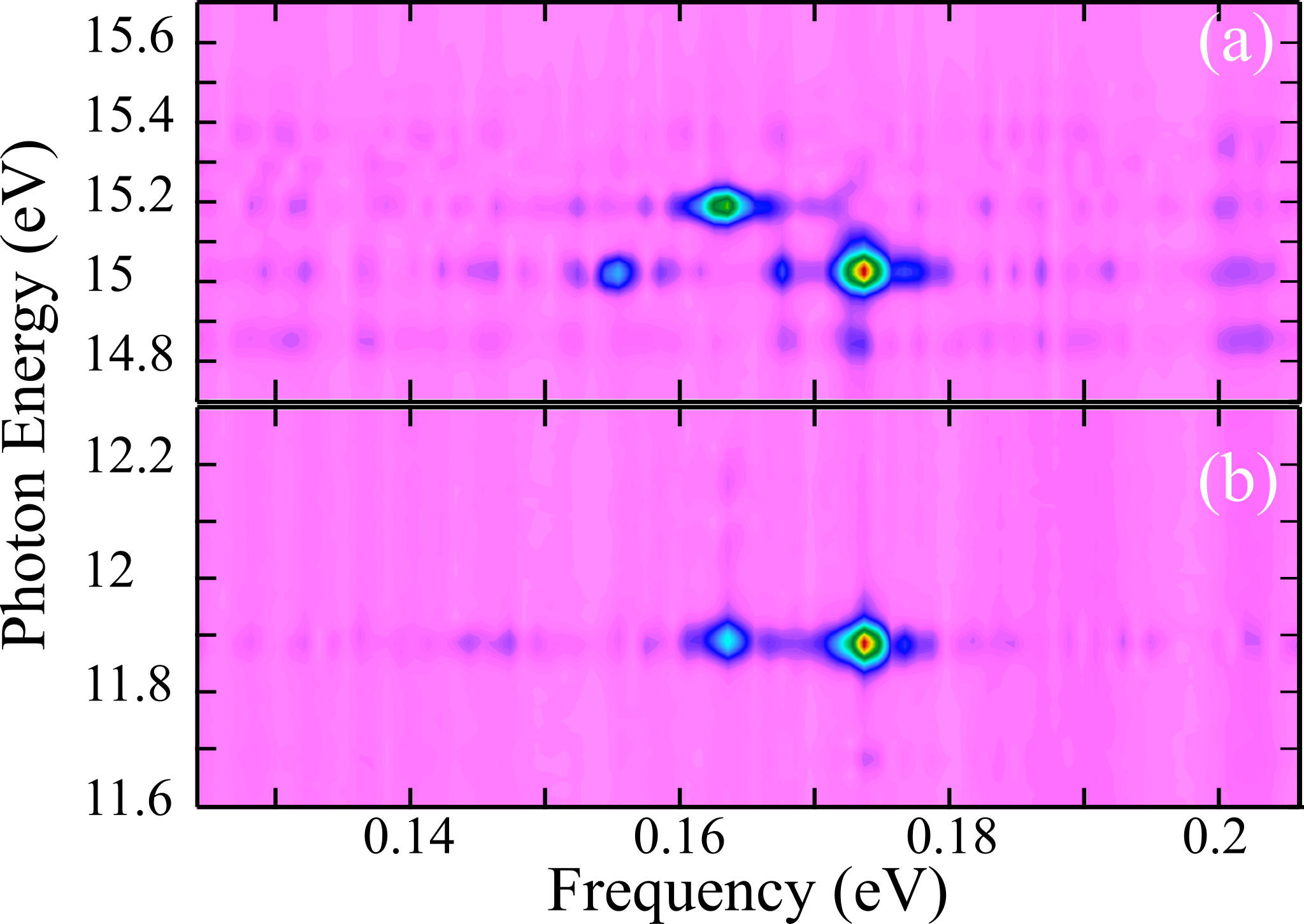}
\caption{High-resolution 2D spectrum for (a) Rydberg states and (b) 4s valence states. We used a parallel relative polarization configuration and the argon backing pressure was 10 mbar.}
\label{fig2}
\end{center}
\end{figure}

\textbf{0-$\pi$ phase plate} 

The 0-$\pi$ phase plate used for the experiments was made by by St\'ephane Guilet at the C2N laboratory in Marcoussis, France. A one inch diameter UV fused silica disk was  patterned using a Capacitively Coupled Plasma Reactive Ion Etching (CCP/RIE) machine through a photo-resist mask defined by UV lithography. Dry etching was carried out by using a low pressure RF plasma of a gas mixture based on $SF_6$ and $CHF_3$. The pressure was around 5 mTorr, the $CHF_3/SF_6$ flow ratio of 2.5 and the RF power around 15 W. This allowed highly anisotropic pattern (angle between walls and the bottom of the etching higher than 88$^{\circ}$). \\

\textbf{Picosecond Time-Resolved XFID} 

We measured the time-dependent ionization damping of the XFID signal to investigate the coherence time of the emission for different argon backing pressure. The XFID signal increases with delay, since the later the IR pulse arrives, the more XUV emission has occured before ionization. 

\hspace{1cm}
\begin{figure}[H]
\begin{center}
\includegraphics[width=0.45\textwidth]{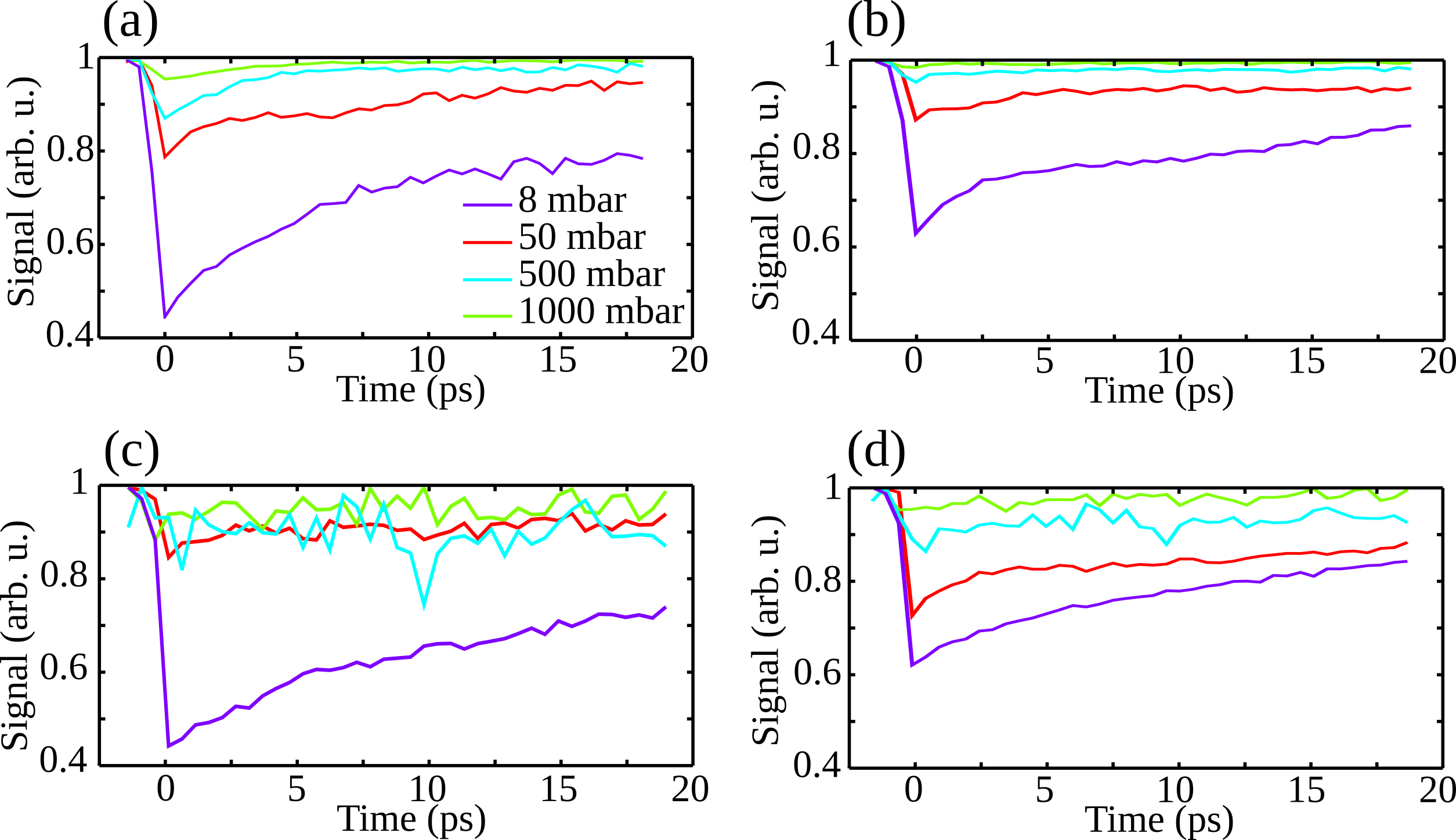}
\caption{Time-resolved XFID for different argon backing pressure. In (a) for R$_1$ (4d/6s ($^2$P$_{3/2}$)), in (b) for R$_2$ (5d/7s ($^2$P$_{3/2}$)), in (c) for R$_3$ (4d/6s ($^2$P$_{1/2}$)) and in (d) for R$_4$ (6d/8s ($^2$P$_{3/2}$)). The signal is normalized to unity (negative pump-probe delay) for each pressure and each Rydberg state.}
\label{fig5}
\end{center}
\end{figure}

Looking at figure \ref{fig5}, one can notice that the caracteristic emission time is strongly shortened when the backing pressure is increases. 

To quantify the caracteristic XFID emission time for the different Rydberg states, we performed an exponantial fitting, for the low-pressure limit (8 mbar argon backing pressure). The fitting provide rising times of 6.3 ps for R$_1$, 30.8 ps for R$_2$, 21.6 ps for R$_3$ and 33.6 ps for R$_4$ (Fig \ref{fig2}). 

\hspace{1cm}
\begin{figure}[H]
\begin{center}
\includegraphics[width=0.45\textwidth]{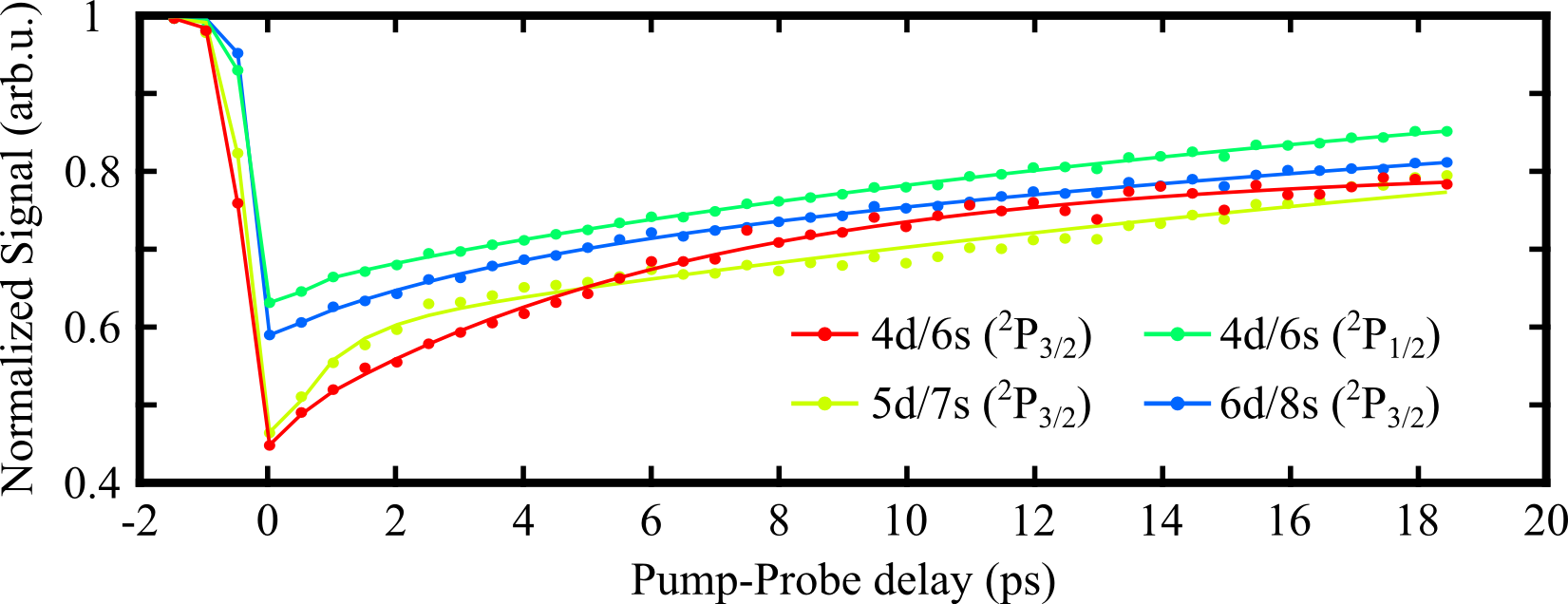}
\caption{Picosecond Time-Resolved XFID for R$_1$ (4d/6s ($^2$P$_{3/2}$)) in red, R$_2$ (5d/7s ($^2$P$_{3/2}$)) in yellow, R$_3$ (4d/6s ($^2$P$_{1/2}$)) in green and for R$_4$ (6d/8s ($^2$P$_{3/2}$)) in blue. The intensity of the IR probe pulse was $\sim 10^{13}$ W/cm$^2$. The argon backing pressure was 8 mbar. The dots are the measurement points and the lines the fits.}
\label{fig2}
\end{center}
\end{figure}

For the fitting of the time-resolved XFID signals, we have used a double-exponantial function. While R$_1$, R$_3$ and R$_4$ are caracterized by only one caracteristic time scale, the decay of the R$_2$ (5d/7s ($^2$P$_{3/2}$) is describe by a long timescale (21.6 ps) and an other much faster timescale (< 2 ps). The origin of this fast timescale is not fully understood yet and would require additional studies. \\

\textbf{Pressure scaling of the signal} 

\begin{figure}[H]
\begin{center}
\includegraphics[width=0.45\textwidth]{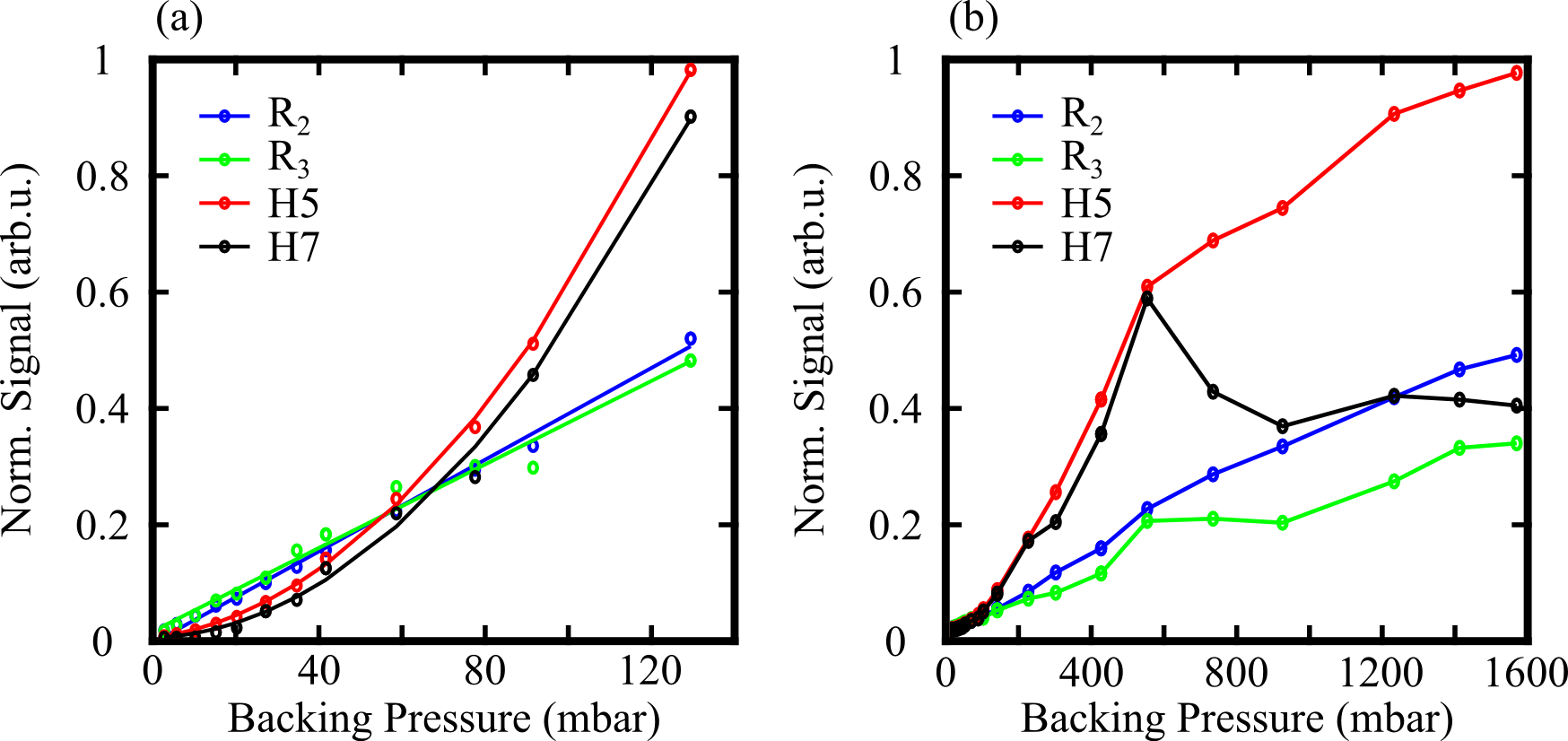}
\caption{Pressure scaling of signals. R$_2$ and R$_3$ are the XFID signals for the 5d/7s($^2$P$_{3/2}$) and 4d/6s($^2$P$_{1/2}$) Rydberg states. H5 and H7 are the below-threshold harmonics 5 and above-threshold harmonic 7. In (a), for the low-pressure range (The dots are experimental data and the lines represent linear fitting for R$_2$, R$_3$ and quadratic fitting H5, H7.) and in (b), up to 1600 mbar argon backing pressure.}
\label{fig4}
\end{center}
\end{figure}

We have studied the pressure scaling of the XFID signal, of the below-threshold harmonic (H5) and of the above threshold harmonic (H7). First, in the low pressure range (0-120 mbar argon backing pressure \ref{fig4}(a)), the harmonics exhibit a quadratic growth as a function fo backing pressure. By contrast the evolution of the XFID signal is remarkably linear. As the backing pressure further increases (Fig. \ref{fig4}(b)), the signal from H7 decreases after 475 mbar and saturates, which is the signature of strong reabsorption. This is not the case for the below-threshold harmonic (H5) and XFID signals (R$_2$ and R$_3$), which continues to grow up to 1600 mbar. 

\hspace{1cm}

Because both HHG and XFID are coherent processes, one would expect a quadratic growth of both signals with atomic density (or backing pressure), at least untill reabsorption effects start playing a role. While this is the case for the HHG signal (H5 and H7), the XFID exibits a linear scaling with pressure. On the other hand, this emission is clearly coherent since it is collimated and produces highly contrasted two-source interference patterns. In order to explain this, we have to take into account the effect of collisions in the experiment.  Because the XFID emission take place on a very long timescale (tens of picosecond), atomic collisions can play a role in the scaling of the signal with pressure. In the next part we derive a simple model that explains the linear pressure dependence of the XFID emission. \\

\textbf{Influence of the collision on the XFID signal} 

\newcommand\partd[2]{\frac{\partial #1}{\partial #2}}
\newcommand\partdd[2]{\frac{\partial^2 #1}{\partial #2^2}}
\renewcommand{\vec}{\overrightarrow}
\renewcommand{\d}{\mathrm{d}}
\newcommand\moyenne[1]{\langle#1\rangle}
\newcommand{\A}{\mathcal{A}}
\newcommand{\K}{k_b}

To simply describe the collision effects in our measurement, let us consider an excited particle that has a given desexcitation radius $r$. If another particle in the ground state is in the sphere of radius $r$, the probability of energy transfer $dP_t$ during $dt$ is considered as constant, and null outside the sphere. Then, with $N_s$ unexcited particles in the sphere :

\begin{equation}
dP_t=B_1 N_s dt
\end{equation}

with $B_1$ a constant independent from the exact inter-particle distance, i.e. from the pressure. If we simply consider $N_s=\frac{PV_{sphere}}{RT}$ then

\begin{equation}
dP_t=B_2 P dt
\end{equation}

Without considering the emission process, the evolution of the excited population $N(t)$ is given by

\begin{equation}
\label{yaya}
dN=-N B_2 P dt \quad \textit{i.e.} \quad \frac{dN}{dt}=-\frac{1}{\tau_{coll}}N \quad with \quad \tau_{coll} \propto \frac{1}{P}
\end{equation}

If we include the emission in equation \eqref{yaya} :

\begin{equation}
\begin{split}
\label{toto}
\frac{dN}{dt}=-\frac{1}{\tau_{em}}N - \frac{1}{\tau_{coll}} N \\
=- N (\frac{1}{\tau_{em}}+\frac{1}{\tau_{coll}}) \\
= -N (\frac{\tau_{em}+\tau_{coll}}{\tau_{em} \tau_{coll}}) 
\end{split}
\end{equation}

The typical timescale of the Rydberg states lifetime is of order of nanoseconds, but our experimental measurements indicate picosecond coherence times (characteristic emission time). This suggests that the collision rate dominates over the emission rates, \textit{i.e.} $\tau_{coll}<<\tau_{em}$ and \eqref{toto} becomes 

\begin{equation}
\frac{dN}{dt}\sim -\frac{1}{\tau_{coll}} N
\end{equation}
Then, 
\begin{equation}
\label{yiyi}
N(t)=N_0 e^{\frac{-t}{\tau_{coll}}}
\end{equation}

Let us express the amount of coherent signal $dS(t)$ emitted at $t$ during $dt$, as the emission only weakly affects the population, as :

\begin{equation}
\label{yoyo}
dS(t)= C_1 N^2(t) dt
\end{equation}

where $C_1$ is a proportionality constant. \\
Integrating \eqref{yoyo} over time with \eqref{yiyi} gives the static XFID signal $S_{tot}$ :

\begin{equation}
S_{tot}= \int^{+\infty}_{0}{dS}= C_1 \int^{+\infty}_{0}{N_0^2 e^{\frac{-2t}{\tau_{coll}}}dt}
\end{equation}
Then, 
\begin{equation}
S_{tot}= C_1 N_0^2 \frac{\tau_{coll}}{2}
\end{equation}

Because $N_0 \propto P$ and according to \eqref{yaya}: $\tau_{coll} \propto \frac{1}{P}$, we obtained : 

\begin{equation}
S_{tot} \propto P
\end{equation}

This shows that even though it is coherent, the total XFID signal increases linearly and not quandratically with pressure because of collisions.

\end{document}